# Light propagation in arbitrary spacetimes and the gravitational lens approximation

Stella Seitz, Peter Schneider & Jürgen Ehlers
Max-Planck-Institut für Astrophysik
Karl-Schwarzschild-Str. 1
Postfach 1523
D–85740 Garching
Federal Republic of Germany

## Abstract

We reformulate the transport equation which determines the size, shape and orientation of infinitesimal light beams in arbitrary spacetimes. The behaviour of such light beams near vertices and conjugate points is investigated, with special attention to the singular behaviour of the optical scalars. We then specialize the general transport equation to the case of an approximate metric of an inhomogeneous universe, which is a Friedmann metric 'on average' with superposed isolated weak matter inhomogeneities. In a series of well-defined approximations, the equations of gravitational lens theory are derived. Finally, we derive a relative optical focusing equation which describes the focusing of light beams relative to the case that the beam is unaffected by matter inhomogeneities in the universe, from which it follows immediately that no beam can be focused less than one which is unaffected by matter clumps, before it propagates through its first conjugate point.

# 1 Introduction

The propagation of light rays in curved spacetimes is described by the equation for null geodesics. Below, we consider congruences of light rays, so-called light beams (for an exact definition, see Sect. 2) and study their propagation in arbitrary spacetimes. Infinitesimal light beams are described by Jacobi's differential equation for deviation vectors. In this paper, we study some properties of the solutions of this propagation equation. In particular, we provide a detailed study of the behaviour of light beams near vertices and conjugate points. The behaviour of the optical scalars (Sachs 1961) which may diverge near conjugate points is determined. We find the leading-order behaviour of the convergence, shear and twist of light beams and their relation to the optical tidal matrix which represents the source of beam deformation.

We then specialize the propagation equation to the case that the metric can be described by that of a Friedmann universe, with superposed weak local inhomogeneities; this is the situation most relevant for the light propagation in the universe. Here, the optical tidal matrix can be split into a contribution due to the background universe and one due to the local inhomogeneities, which is described in the first post-Minkowskian approximation. The background universe is assumed to have the overall geometry of a smooth Friedmann universe, but is locally modified due to matter inhomogeneities.

If the matter inhomogeneities along the light beam are well localized, i.e., if the spatial extent of the inhomogeneities is much smaller than the distance from the source to an observer, the contributions from the inhomogeneities can be described in the impulse approximation, in which the contribution to the optical tidal matrix due to inhomogeneities is replaced by a sum of delta-distributions. We will then show that this approximation leads to the gravitational lens equations, which are usually used to describe the influence of weak matter inhomogeneities on light propagation in the universe (for a review on gravitational lens theory and its applications, see Schneider, Ehlers & Falco 1992, hereafter SEF). Hence, the gravitational lens equations follow from the exact propagation equations for light beams with a series of well-defined approximations.

The behaviour of the cross-sectional area of an infinitesimally small light beam is described by the optical focusing equation (Sachs 1961), which contains the trace of the optical tidal matrix and the shear of the light beam as source terms. We will show that a *relative optical focusing equation* can be obtained which describes the cross-sectional area of a beam relative to one which is unaffected by matter inhomogeneities. The uniquely-determined independent variable for this relative focusing equation is the $\chi$-function introduced for other reasons in Sect. 4.6 of SEF. From this relative focusing equation it follows directly that no light beam can be less focused than one which is unaffected by matter inhomogeneities before the beam propagates through its first conjugate point. In the frame of gravitational lens theory, this fact has been proved earlier (Schneider 1984, Seitz & Schneider 1992, hereafter Paper I, 1994).

# 2 Infinitesimal light beams

In this section we review some consequences of the fact that, according to the geometrical optics approximation to Maxwell's equations in an arbitrary spactime $(M, g_{\alpha\beta})$, a locally nearly plane electromagnetic wave, propagating without interaction with matter, is associated with a hypersurface-orthogonal congruence of null geodesics representing *light rays*. We denote the corresponding phase function by $S$ and the wave vector by $k^\alpha = -g^{\alpha\beta} S_{,\beta}$; then $k_\alpha k^\alpha = 0$ and $\dot{k}^\alpha := k_{\alpha;\beta} k^\beta = 0$. (For details concerning this section see, e.g., SEF, Chapt. 3 and Wald 1984, Sect. 9.2 & 9.3., see also Blandford & Narayan 1992.)

We fix attention on one "central" light ray $\gamma_0$ and denote by $Y^\alpha$ any *deviation vector field* (Jacobi field) "connecting" $\gamma_0$ to one of its neighbours. Then, $k_\alpha Y^\alpha$ is constant on $\gamma_0$. Deviation vectors differing by a (constant) multiple of $k^\alpha$ represent displacements to the same nearby ray. Given the four velocity $U^\alpha$ of an observer at an event $p$ on $\gamma_0$, one can always arrange that $Y^\alpha$ is spatial for $U^\alpha$, i.e., $U_\alpha Y^\alpha = 0$.

Two events $p$, $q$ on $\gamma_0$ are said to be *conjugate* if there exists a not identically vanishing Jacobi field which is zero at $p$ and $q$. For such a Jacobi field, $k_\alpha Y^\alpha = 0$. A deviation vector satisfying the last equation (whether it vanishes somewhere or not) connects rays contained in the same phase hypersurface $S = $ const.

Henceforth we consider exclusively 2-parameter families of rays contained in one phase hypersurface which we call *beams*. Their deviation vectors obey $k_\alpha Y^\alpha = 0$, consequently the size, shape and orientation of an infinitesimal cross section of a beam is independent of the 4-velocity of the oberver who measures it.

Given the 4-velocity $U^\alpha$ of an observer at an event $p$ on $\gamma_0$, one can choose deviation vectors to all neighbouring rays such that, besides $k_\alpha Y^\alpha = 0$, also $U_\alpha Y^\alpha = 0$. Such vectors $Y^\alpha$ span a 2-dimensional, spacelike subspace of the tangent space $M_p$ of $p$ which we call a *screen* adapted to $k^\alpha$, $U^\alpha$.

In studying conjugate pairs on a ray $\gamma_0$ it suffices to consider deviation vectors belonging to a beam surrounding $\gamma_0$.

For gravitational lensing, the important beams are those which are contained in either the future *null cone* $\mathcal{C}_s^+$ of an event $s$ – flashes of light emitted from a source event $s$ – or the past null cone $\mathcal{C}_o^-$ of an observation event $o$. (In the second case, the rays of a beam belong to different, usually mutually incoherent locally plane waves, emitted from different source events. This does not matter for the applications considered in this paper. It is often helpful to think of the rays as [classical models of] photons.) In the remainder of this paper we are concerned with such beams only.

$\mathcal{C}_o^-$ is generated by all null geodesic rays ending at $o$. The set of all events conjugate to the vertex $o$ on those rays forms the *caustic* of $\mathcal{C}_o^-$. $\mathcal{C}_o^-$ has the shape of a (hyper-) cone only between $o$ and the first sheet of the caustic; thereafter in general it bifurcates and intersects itself. This is the (theoretical) reason for the phenomenon of multiple imaging in gravitational lensing.

Consider an observer at the event $o$ with 4-velocity $U_o^\alpha$, $U_o^\alpha U_{o\alpha} = 1$, and the past light cone $\mathcal{C}_o^-$. Choose the affine parameter $\lambda$ of the rays ending at $o$ such that *(i)* $\lambda = 0$ at $o$, *(ii)* $\lambda$ increases to the past, *(iii)* at $o$, $\tilde{k}_\alpha U_o^\alpha = -1$. Then, $\tilde{k}^\alpha = \frac{dx^\alpha}{d\lambda}$ is past-directed, and for events on $\mathcal{C}_o^-$ infinitely close to $o$, $d\lambda$ is the distance from $o$ measured by the chosen observer. The "new" $\tilde{k}^\alpha$ is related to the wave vector introduced above by $k^\alpha = -\frac{\omega_o}{c} \tilde{k}^\alpha$ if $\omega_o$ is the frequency associated with $k^\alpha$ at the observer. $\tilde{k}^\alpha$ is purely kinematical, the same for all monochromatic waves which might be travelling in the direction $-\tilde{k}^\alpha$. Let $\gamma_0$ be a ray, and let $U^\alpha$ on $\gamma_0$ be the result of parallelly propagating $U_o^\alpha$. Choose, along $\gamma_0$, orthonormal bases $(E_1^\alpha, E_2^\alpha)$ on the screens adapted to $\tilde{k}^\alpha$, $U^\alpha$, parallel on $\gamma_0$. The deviation vectors of the beam centered on $\gamma_0$ can then be written as $Y^\alpha = -\xi_1 E_1^\alpha - \xi_2 E_2^\alpha - \xi_o \tilde{k}^\alpha$; then the screen components $\xi_i$ ($i = 1, 2$) change according to the deformation equation

$$\dot{\xi}_i = S_{ij}\xi_j \quad , \quad S_{ij} = E_i^\alpha \tilde{k}_{\alpha;\beta} E_j^\beta \quad ,$$

where a dot denotes differentiation with respect to the affine parameter. In matrix notation we write

$$\dot{\boldsymbol{\xi}} = \mathcal{S}\boldsymbol{\xi} \qquad (2.1a)$$

The *optical deformation matrix* $\mathcal{S}$ is composed of Sachs' optical scalars of the beam (Sachs 1961), i.e., its rate of expansion

$$\theta(\lambda) := \frac{1}{2}\tilde{k}^\alpha_{;\alpha}(\lambda)$$

and its (complex) rate of shear,

$$\sigma(\lambda) := \frac{1}{2}\tilde{k}_{\alpha;\beta}(\lambda)\,\epsilon^{*\alpha}(\lambda)\,\epsilon^{*\beta}(\lambda) \quad , \quad \epsilon^\alpha := E_1^\alpha + iE_2^\alpha \quad ,$$

according to

$$\mathcal{S}(\lambda) := \begin{pmatrix} [\theta(\lambda) - \mathcal{R}e\sigma(\lambda)] & [\mathcal{I}m\sigma(\lambda)] \\ [\mathcal{I}m\sigma(\lambda)] & [\theta(\lambda) + \mathcal{R}e\sigma(\lambda)] \end{pmatrix} \quad . \qquad (2.1b)$$

Since $k_\alpha = -S_{,\alpha}$ is the gradient of the phase $S$, $\tilde{k}_{\alpha;\beta} = \tilde{k}_{\beta;\alpha}$, and therefore $\mathcal{S}$ is a symmetric matrix. Differentiation of (2.1a) with respect to $\lambda$ gives

$$\ddot{\boldsymbol{\xi}}(\lambda) = \mathcal{T}(\lambda)\boldsymbol{\xi}(\lambda) \quad . \qquad (2.2a)$$

where

$$\dot{\mathcal{S}} + \mathcal{S}^2 = \mathcal{T} \quad . \qquad (2.2b)$$

Combining the last equation with Sachs' transport equations for $\theta$ and $\sigma$,

$$\dot{\theta} + \theta^2 + |\sigma|^2 = \mathcal{R} \quad , \qquad (2.2c)$$

$$\dot{\sigma} + 2\theta\sigma = \mathcal{F} \quad , \qquad (2.2d)$$

shows that the *optical tidal matrix* $\mathcal{T}$ is given by

$$\mathcal{T}(\lambda) := \begin{pmatrix} [\mathcal{R}(\lambda) - \mathcal{R}e\mathcal{F}(\lambda)] & [\mathcal{I}m\mathcal{F}(\lambda)] \\ [\mathcal{I}m\mathcal{F}(\lambda)] & [\mathcal{R}(\lambda) + \mathcal{R}e\mathcal{F}(\lambda)] \end{pmatrix} \quad , \qquad (2.2e)$$

where

$$\mathcal{R}(\lambda) := -\frac{1}{2}R_{\beta\gamma}(\lambda)\tilde{k}^\beta(\lambda)\tilde{k}^\gamma(\lambda) \quad , \qquad (2.2f)$$

$$\mathcal{F}(\lambda) := -\frac{1}{2}C_{\alpha\beta\gamma\delta}(\lambda)\,\epsilon^{\alpha*}(\lambda)\tilde{k}^\beta(\lambda)\epsilon^{\gamma*}(\lambda)\tilde{k}^\delta(\lambda) \quad . \qquad (2.2g)$$

Similar equations have been derived by Blandford et al. (1991) and Peebles (1993, Chapt. 14).[1] The optical tidal matrix is symmetric due to the symmetry $C_{\alpha\beta\gamma\delta} = C_{\delta\gamma\beta\alpha}$ of the conformal curvature tensor. Equations (2.2a,e,f&g) exhibit how the Ricci and conformal curvatures govern the evolution of infinitesimal light beams; they are equivalent to the *geodesic deviation equation* (Jacobi equation) *for screen vectors.*[2]

---

[1] Note, however, that the component $\propto k^\alpha$ of $Y^\alpha$ cannot be made to vanish for all $\lambda$, contrary to the claim in Pebbles (consider equation (14.9) in his book, where $\xi^i$ corresponds to 'our' $Y^\alpha$.)

The linearity of the Jacobi equation (2.2a) implies that the solution $\boldsymbol{\xi}(\lambda)$ is related to its initial value $\dot{\boldsymbol{\xi}}(0) =: \boldsymbol{\theta}$ by a $\lambda$-dependent linear transformation

$$\boldsymbol{\xi}(\lambda) = \mathcal{D}(\lambda)\boldsymbol{\theta} \quad . \tag{2.3}$$

With the choice of $\lambda$ described above, $\boldsymbol{\theta}$ is the (vectorial) angle between $\gamma_0$ and a neighbouring ray. Because of (2.2a), $\boldsymbol{\xi}(0) = 0$ and $\dot{\boldsymbol{\xi}}(0) = \boldsymbol{\theta}$, $\mathcal{D}(\lambda)$ is determined by

$$\ddot{\mathcal{D}}(\lambda) = \mathcal{T}(\lambda)\mathcal{D}(\lambda) \quad , \tag{2.4a}$$

$$\mathcal{D}(0) = \mathcal{O} \quad , \quad \dot{\mathcal{D}}(0) = \mathcal{I} \tag{2.4b}$$

or, equivalently, by the linear integral equation

$$\mathcal{D}(\lambda) = \lambda\mathcal{I} + \int_0^\lambda d\lambda' \, (\lambda - \lambda') \, \mathcal{T}(\lambda')\mathcal{D}(\lambda') \quad . \tag{2.5}$$

The *Jacobi map* (2.3) takes infinitesimal changes of ray directions at the observer back to a screen at an event of $\gamma_0$ given by the value of $\lambda$. If that event is taken on some source "plane" $z = \text{const}$, $\mathcal{D}(\lambda)$ corresponds to the properly scaled *magnification matrix* (in the terminology of SEF) of lens theory. Note that in contrast to $\mathcal{S}$ and $\mathcal{T}$, $\mathcal{D}$ is in general not symmetric.

Equation (2.4a) implies:

1) If $\mathcal{T}(\lambda)$ is continuously differentiable $k$ times, $\mathcal{D}(\lambda)$ is continuously differentiable $k+2$ times; assuming $k$ sufficiently large (which is permissable) justifies our later use of Taylor polynomials to study the local properties of $\mathcal{D}(\lambda)$ at special points.

2) $\dot{\mathcal{D}}^T\mathcal{D} - \mathcal{D}^T\dot{\mathcal{D}}$ is a first integral of (2.4a). Since it vanishes in consequence of the initial conditions (2.4b), *all solutions of (2.4) obey $\dot{\mathcal{D}}^T\mathcal{D} = \mathcal{D}^T\dot{\mathcal{D}}$, provided $\mathcal{T}$ is continuous there. At discontinuities and $\delta$-type singularities of $\mathcal{T}$ this relation is preserved.*

According to the definitions given above, $\lambda_c$ corresponds to a point $p_c$ conjugate to the vertex (observer) if and only if $\det \mathcal{D}(\lambda_c) = 0$. If the rank of $\mathcal{D}(\lambda_c)$ is equal to zero, i.e. if $\mathcal{D}(\lambda_c) = \mathcal{O}$, all rays arriving at $o$ have been intersecting to first order at $p_c$; if the rank of $\mathcal{D}(\lambda_c)$ is equal to one, the cross section of the ray bundle has been degenerating into an infinitesimal line segment at $p_c$. In the first case, $p_c$ is called a *focus* (or degenerate conjugate point) of the caustic of $\mathcal{C}_0^-$, in the second case, it is said to be a non-degenerate or simple conjugate point.

Comparison of (2.1a) and the derivative of (2.3) shows that

$$\dot{\mathcal{D}} = \mathcal{S}\mathcal{D} \quad ; \tag{2.6}$$

thus $\mathcal{S}$ can be obtained from $\mathcal{D}$ (see below, Sect. 3). With (2.6) we alternatively derive the symmetry of the $\mathcal{S}$-matrix from the 'basic' differential equation (2.4a) and the vertex-initial conditions (2.4b): at an affine parameter where $\mathcal{D}^{-1}$ and thus $\mathcal{S}$ exist, $\dot{\mathcal{D}}^T\mathcal{D} = \mathcal{D}^T\dot{\mathcal{D}}$ is equivalent to $\mathcal{S} = \mathcal{S}^T$. This also implies that at points where $\det \mathcal{D} = 0$, the antisymmetric part of $\mathcal{S}$ is equal to zero.

---

[2] In equation (2.2g) one may write the full curvature tensor instead of $C_{\alpha\beta\gamma\delta}$; the Ricci part does not contribute.

Consider now the determinant of the Jacobi map. From its definition in (2.3) it follows that its absolute value is equal to the area $\delta A(\lambda)$ of the cross section of the light beam at this affine parameter, divided by the solid angle $\delta\Omega$ which that cross section subtends at the observer:

$$|\det \mathcal{D}(\lambda)| = \frac{\delta A(\lambda)}{\delta \Omega} \quad . \tag{2.7}$$

At a non-degenerate conjugate point the Jacobian determinant changes its sign; at a focus its sign is conserved, as will be shown in Sect. 3. Thus, $\det \mathcal{D}(\lambda)$ contains information about the area $\delta A(\lambda)$ as well as the parity, i.e., the orientation of a beam at $\lambda$ relative to that close to the vertex. Between the vertex and conjugate points, the area $\delta A(\lambda)$ is governed by Sachs' focusing equation:

$$\left(\sqrt{A(\lambda)}\right)^{\cdot\cdot} = \left[\mathcal{R}(\lambda) - |\sigma(\lambda)|^2\right] \sqrt{A(\lambda)} \quad . \tag{2.8a}$$

This ordinary differential equation has $\mathcal{C}^2$-solutions in any $\lambda$-interval in which $\mathcal{R}(\lambda) - |\sigma(\lambda)|^2$ is continuous. This is the case except if the interval contains simple conjugate points, see Sect.3. The initial conditions for the solution of (2.8a), which gives the area of the beam, are $\sqrt{A(0)} = 0$ and $\frac{d\sqrt{A}}{d\lambda}(0) = \Omega$, where $\Omega$ is the solid angle of the beam at the observer. If there is an odd number of nondegenerate conjugate points between the observer and $\lambda$, one has to take the negative root of $A$, otherwise the positive one. The driving term of the focusing equation, $\mathcal{R} - |\sigma|^2$, is nonpositive: the Einstein field equation with an energy momentum-tensor of an ideal fluid yields a non-positive *source of convergence* $\mathcal{R}$; this also holds for a cosmological constant. Hence, equation (2.8a) describes how a light beam is focused at $\lambda$ due to the "local" curvature (Ricci-focusing) and due to its own shear rate at this affine parameter. Since this shear rate was produced by the *source of shear* $\mathcal{F}$ at a smaller $\lambda$, this implies that both, $\mathcal{R}$ and $\mathcal{F}$, yield a focusing of the light beam. Hence, as long as one considers only the area and not the shape of a light beam, the actions of $\mathcal{R}$ and $\mathcal{F}$ are not distinguishable. In the following we do not consider the evolution of the area of a light beam, but that of

$$w(\lambda) := \mathcal{SQ}[\det \mathcal{D}](\lambda) \equiv \text{sign}\left(\det \mathcal{D}(\lambda)\right) \sqrt{|\det \mathcal{D}(\lambda)|} \quad ; \tag{2.9}$$

the absolute value, $|w|(\lambda) = \sqrt{|\det \mathcal{D}(\lambda)|} = \sqrt{\frac{\delta A(\lambda)}{\delta \Omega}}$, of this function describes the angular diameter distance along the beam considered, and the sign is the parity of the Jacobi map. From (2.7) we obtain that $w$ also fullfills the focusing equation

$$\ddot{w}(\lambda) = \left[\mathcal{R}(\lambda) - |\sigma(\lambda)|^2\right] w(\lambda) \tag{2.8b}$$

between conjugate points; the initial conditions for $w$ are: $w(0) = 0$ and $\dot{w}(0) = 1$. It is not clear a priori how to connect the solutions between conjugate points with each other, or whether one at all can integrate over conjugate points: the matrix $\mathcal{S}$ of eq. (2.6) and thus $\theta$ and $\sigma$ become singular at the vertex and at a conjugate point $\lambda_c$. We investigate the behaviour of a light beam near the vertex and a conjugate point in the next Section and show that the solution of (2.8) is nevertheless well defined at conjugate points between source and observer.

# 3 The behaviour of light beams near vertices and conjugate points

**Preliminaries: Parametrization of a $2 \times 2$-matrix**
For our further discussion we parametrize a real $2 \times 2$-matrix $A$ in terms of [3] 'convergence' $\tau$, 'twist' $\omega$ and 'shear' $\Gamma_1$ and $\Gamma_2$ and write them as real and imaginary parts of complex numbers $\Lambda$ and $\Gamma$, respectively:

$$\tau[A] := \frac{1}{2}(a_{11} + a_{22}) \quad , \quad \omega[A] := \frac{1}{2}(a_{12} - a_{21}) \quad , \quad \Lambda[A] = \tau[A] + i\omega[A] \quad , \quad (3.1)$$

$$\Gamma_1[A] := \frac{1}{2}(a_{11} - a_{22}) \quad , \quad \Gamma_2[A] := \frac{1}{2}(a_{12} + a_{21}) \quad , \quad \Gamma[A] = \Gamma_1[A] + i\Gamma_2[A] \quad . \tag{3.2}$$

Then, the trace of $A$ is $\text{tr}A = 2\mathcal{R}e\Lambda[A]$ and its determinant is $\det A = |\Lambda[A]|^2 - |\Gamma[A]|^2$. Note that transforming $A$ with a proper orthogonal matrix (rotation matrix)

$$R(\vartheta) = \begin{pmatrix} \cos\vartheta & \sin\vartheta \\ -\sin\vartheta & \cos\vartheta \end{pmatrix}$$

to $A' = R^{-1}AR$ leaves $\Lambda$ invariant ($\Lambda[A'] = \Lambda[A]$) and transforms $\Gamma$ to $\Gamma[A'] = \Gamma[A]e^{2i\vartheta}$. $\Lambda$ and $|\Gamma|$ have an intrinsic, coordinate-independent meaning for the map given by $A$, whereas the phase of $\Gamma$ fixes the coordinate-system to which $A$ refers. We illustrate our definitions for $\mathcal{S}$ and $\mathcal{T}$:

$$\Lambda[\mathcal{S}](\lambda) = \theta(\lambda) \in \mathbb{R} \quad , \quad \Gamma[\mathcal{S}](\lambda) = -\sigma^*(\lambda) \quad , \tag{3.3}$$

$$\Lambda[\mathcal{T}](\lambda) = \mathcal{R}(\lambda) \in \mathbb{R} \quad , \quad \Gamma[\mathcal{T}](\lambda) = -\mathcal{F}^*(\lambda) \quad . \tag{3.4}$$

If the argument of $\Lambda$ is the Jacobian matrix $\mathcal{D}$, we simply write $\Lambda[\mathcal{D}] =: \Lambda$, and obtain for the derivatives with respect to $\lambda$ that $\Lambda\left[\dot{\mathcal{D}}\right](\lambda) = \dot{\Lambda}(\lambda)$ and $\Lambda\left[\ddot{\mathcal{D}}\right](\lambda) = \ddot{\Lambda}(\lambda)$; analogous relations hold for $\Gamma$. This complex formalism is very convenient for matrix operations; e.g., we obtain for the multiplication of real $2 \times 2$-matrices $A$ and $B$

$$\begin{aligned}\Lambda[AB] &= \Lambda[A]\Lambda[B] + \Gamma^*[A]\Gamma[B] \quad , \\ \Gamma[AB] &= \Gamma[A]\Lambda[B] + \Lambda^*[A]\Gamma[B] \quad .\end{aligned} \tag{3.5}$$

To obtain the geometrical interpretation of $\Lambda$ and $\Gamma$, we consider the polar decomposition of $\mathcal{D}$. If $\mathcal{D} \neq \mathcal{O}$, there exist unique numbers $b_1$, $b_2$, with $0 < b_1 \geq b_2$, and unique angles $\phi$ and $\vartheta$, $0 \leq \phi \leq \pi$, $0 \leq \vartheta \leq 2\pi$, such that

$$\mathcal{D} = R(\vartheta)B(b_1, b_2, \phi) \quad ;$$

$R$ is the rotation matrix which was already defined, and $B$ is a symmetric matrix $B = B^T$:

$$B = R(-\phi)\begin{bmatrix} b_1 & 0 \\ 0 & b_2 \end{bmatrix}R(\phi) \quad .$$

---

[3]     These names are chosen for convenience and are not intended to contain a geometrical meaning. In the case of the Jacobi matrix, the geometrical interpretation of $\Lambda$ and $\Gamma$ will be given below eq. (3.5).

In the polar decomposition $b_1$, $b_2$ and $\vartheta$ are the coordinate invariant numbers, $\phi$ depends on the chosen coordinate system. The matrix $B$ describes a rotation-free deformation, whereas $R(\vartheta)$ rotates the plane by an angle $\vartheta$. The relation of $\{\Lambda, \Gamma\}$ to $\{b_1, b_2, \vartheta, \phi\}$ can be derived with (3.5); we obtain:

$$|\Lambda| \pm |\Gamma| = b_{1,2} \quad , \quad \Lambda = \frac{1}{2}(b_1 + b_2)\,\mathrm{e}^{\mathrm{i}\vartheta} \quad ,$$

$$\Gamma = \frac{1}{2}(b_1 - b_2)\,\mathrm{e}^{\mathrm{i}(2\phi - \vartheta)} \quad .$$

Inserting the values of $b_1$ and $b_2$ yields that the 'twist' $\omega$ is related to the rotation angle $\vartheta$ of the Jacobi map via

$$\frac{\Lambda}{|\Lambda|} = \mathrm{e}^{\mathrm{i}\vartheta} \quad \Rightarrow \quad \tan\vartheta = \frac{\omega}{\tau} \quad . \tag{3.6}$$

## 3.1 Consequences of the symmetry of $\mathcal{S}$

1) Below (2.5) we have derived that $\dot{\mathcal{D}}^T \mathcal{D} = \mathcal{D}^T \dot{\mathcal{D}}$, or $\omega\left[\dot{\mathcal{D}}^T \mathcal{D}\right] = 0$ or, that $\mathcal{S}$ is symmetric. Evaluating the twist part of $\dot{\mathcal{D}}^T \mathcal{D}$ yields:

$$\mathcal{I}\mathrm{m}\left\{\dot{\Lambda}^* \Lambda + \dot{\Gamma}^* \Gamma\right\} = 0 \quad . \tag{3.7}$$

This constraint-equation is valid at every affine parameter and in particular at the vertex and at every conjugate point. Equation (3.7) illustrates that solving for $\ddot{\mathcal{D}} = \mathcal{T}\mathcal{D}$, one has not 8 but only 7 free initial conditions. If one chooses the alternative way to solve for the light propagation – evaluating the optical scalars and than solving $\dot{\mathcal{D}} = \mathcal{S}\mathcal{D}$ – then one has a priori only 7 free initial conditions and the constraint equation (3.7) is hidden in the nonlinear differential equations for the optical scalars.

2) Consider a light beam in an intervall $\lambda \in [\lambda_n, \lambda_{n+1}]$ where $\mathcal{T}(\lambda) = \mathcal{R}(\lambda)\mathcal{I}$, i.e. where the source of shear vanishes. Then every component of $\mathcal{D}$ satisfies the same differential equation, and the general solution $\mathcal{D}$ is a linear combination of two linearly independent solutions $f$ and $g$ of $\ddot{x} = \mathcal{R}x$:

$$\mathcal{D}(\lambda) = f(\lambda)\mathcal{D}_n + g(\lambda)\mathcal{D}_{n+1} \quad , \tag{3.8}$$

where $\mathcal{D}(\lambda_i) =: \mathcal{D}_i$ and $f_n = 1$, $f_{n+1} = 0$, $g_n = 0$ and $g_{n+1} = 1$.[4] Since $g$ and $f$ are linearly independent solutions, we also have $\dot{g}_n \neq 0$ and $\dot{f}_{n+1} \neq 0$. Inserting (3.8) into $\dot{\mathcal{D}}^T \mathcal{D} = \mathcal{D}^T \dot{\mathcal{D}}$ and evaluating this matrix at $\lambda_n$ yields $\mathcal{D}_{n+1}^T \mathcal{D}_n = \mathcal{D}_n^T \mathcal{D}_{n+1}$. Hence we have shown: if there is no source of shear between $\lambda_n$ and $\lambda_{n+1}$, the matrix product

$$\mathcal{D}_{n+1}^T \mathcal{D}_n \tag{3.9a}$$

is symmetric. If one matrix (say $\mathcal{D}_n$) is not singular, i.e., there is no (to the vertex) conjugate point at $\lambda_n$, then the symmetry of (3.9a) can be expressed as the statement that the matrix

$$\mathcal{D}_{n+1}\mathcal{D}_n^{-1} \tag{3.9b}$$

---

[4] This is a Sturm boundary value problem: the functions $f$ and $g$ exist if and only if the solution with $x_n = 0$ and $\dot{x}_n = 1$ satisfies $x_{n+1} \neq 0$. This condition is violated if and only if $\lambda_{n+1}$ and $\lambda_n$ correspond to a pair of conjugate points.

which carries connection vectors from $\lambda_n$ to $\lambda_{n+1}$ is symmetric. This property has been used extensively in the proof of the magnification theorem in gravitational lens theory in Paper I.

### 3.2 The Jacobi map near a vertex

At the vertex, $\Lambda = 0 = \Gamma$, $\dot{\Gamma} = 0 = \mathcal{I}\mathrm{m}\dot{\Lambda}$ and $\mathcal{R}\mathrm{e}\dot{\Lambda} = 1$. In this Section we do not investigate the behaviour of the optical scalars at the vertex since it is the same as that near a focus; this is due to the fact that locally a beam at a focus differs from that at a vertex only by the opening angle; this angle cancels out in the optical scalars because they are relative quantities. We investigate the Jacobi map in a Taylor expansion as a function of $\epsilon = \lambda - \lambda_\mathrm{v} = \lambda$; we put $\mathcal{T}(0) =: \mathcal{T}_0$ and obtain with (2.4)

$$\mathcal{D}(\epsilon) = \mathcal{I}\epsilon + \frac{1}{6}\mathcal{T}_0\epsilon^3 + \mathcal{O}(\epsilon^4) \quad . \tag{3.10}$$

Eq. (3.10) implies with the symmetry of $\mathcal{T}_0$ that the shear of the Jacobi map is at least of third order near the vertex, and the twist increases even slower at the vertex. In other words, the cross section of an initially circular light beam becomes distorted to an ellipse before it can get twisted. To compare the evolution of the shear of the Jacobian with its twist in more detail, we claim: *if the first nonvanishing contribution to $\Gamma$ is of the order $\epsilon^n$, $n \geq 3$, at a vertex, the leading term of $\omega$ is at least of the order $\epsilon^{2n}$* (generically, $n = 3$).

For the proof, we insert the Taylor expansions of $\Gamma_1$ and $\Gamma_2$ into the constraint equation (3.7); this yields that the first nonvanishing contribution of this term is of the order $2n$. Inserting the Taylor expansions of $\tau$ and $\omega$ and using that the first nonvanishing contribution to $\tau$ is of order one we find that, in order to satisfy the constraint equation at every order of $\epsilon$, the leading order of $\omega$ must be at least $2n$.

Therefore, the twist $\omega$ increases at the vertex very slowly compared to the shear; this explains that "not too far" from the observer, the light beam can not be twist-dominated, i.e. $\omega^2 < |\Gamma|^2$ holds. This slow increase also holds for the rotation-angle $\vartheta$ of the polar decomposition of the Jacobian matrix near the vertex, since with (3.6) $\tan\vartheta = \frac{\omega}{\tau}$. With $\tau(\epsilon) = \epsilon + \mathcal{O}(\epsilon^3)$ and $\omega = a\epsilon^6 + \mathcal{O}(\epsilon^7)$, the rotation angle $\vartheta = \arctan\frac{\omega}{\tau}$ becomes near the vertex $\vartheta(\epsilon) = a\epsilon^5 + \mathcal{O}(\epsilon^6)$.

### 3.3 The light beam near a conjugate point

Non-degenerate conjugate points $\lambda_\mathrm{c}$ are characterized by $0 \neq |\Gamma(\lambda_\mathrm{c})| = |\Lambda(\lambda_\mathrm{c})|$. Since $\Lambda$, but not $\Gamma$ is invariant under rotation of the coordinate system, we can orient the latter such that $\Gamma(\lambda_\mathrm{c}) = \Lambda(\lambda_\mathrm{c})$ at the conjugate point. At a focus, $\Gamma(\lambda_\mathrm{c}) = 0 = \Lambda(\lambda_\mathrm{c})$. In the following we describe the light beam, as before near the vertex, in a Taylor-expansion around the conjugate point as a function of $\epsilon := \lambda - \lambda_\mathrm{c}$. We first derive properties which are common to both kinds of conjugate points; investigating the local behaviour of beams at a conjugate point, we are only interested in solutions of (2.4a) which obey the initial conditions (2.4b).

Theorem: *At a conjugate point $x_\mathrm{c}$ an eigenvector belonging to the eigenvalue zero of $\mathcal{D}_\mathrm{c}$ cannot be a zero eigenvector to $\dot{\mathcal{D}}_\mathrm{c}$. In particular, this implies that at a focus the rank of $\dot{\mathcal{D}}_\mathrm{c}$ is two, and at a non-degenerate conjugate point the rank of $\dot{\mathcal{D}}_\mathrm{c}$ is at least equal to one.*

Proof: Assume that there exists a conjugate point where $\mathcal{D}_c \mathbf{x} = \mathbf{0}$ and $\dot{\mathcal{D}}_c \mathbf{x} = \mathbf{0}$. Let $\boldsymbol{\xi}(\lambda) = \mathcal{D}(\lambda)\mathbf{x}$. Then this Jacobi field obeys $\boldsymbol{\xi}_c = \mathbf{0}$, $\dot{\boldsymbol{\xi}}_c = \mathbf{0}$, and hence $\boldsymbol{\xi} \equiv 0$ and also $\dot{\boldsymbol{\xi}}(0) = \mathbf{0}$ which is in contradiction to $\dot{\boldsymbol{\xi}}(0) = \boldsymbol{\theta} \neq \mathbf{0}$. q.e.d.

In order to derive Taylor expansions of $\det \mathcal{D}$, $\theta$ and $\sigma$ near conjugate points, we consider the differential equation $\ddot{\mathcal{D}} = \mathcal{T}\mathcal{D}$. Using (3.5) we rewrite this linear matrix differential equation as system of coupled differential equations for $\Lambda$ and $\Gamma$:

$$\ddot{\Lambda} - \mathcal{R}\Lambda = -\mathcal{F}\Gamma \quad , \quad \ddot{\Gamma} - \mathcal{R}\Gamma = -\mathcal{F}^*\Lambda \quad , \tag{3.11}$$

which describe two coupled, planar oscillators with the same eigen-frequency and the same absolute coupling strength. Taking the $n$-th derivative of (3.11), one can iteratively calculate the Taylor-expansion coefficients of $\Gamma$ and $\Lambda$ in the $(n+2)$-th order as a function of $\Lambda_c$, $\Gamma_c$, $\dot{\Lambda}_c$ and $\dot{\Gamma}_c$ (for the case of a non-degenerate conjugate point) or as a function of $\dot{\Lambda}_c$ and $\dot{\Gamma}_c$ (for the case of a focus). A conserved quantity of the differential equation system (3.11) is

$$\mathcal{L} := \dot{\Lambda}^* \Lambda - \Lambda \dot{\Lambda}^* + \Gamma^* \dot{\Gamma} - \Gamma \dot{\Gamma}^* \quad . \tag{3.12}$$

Thus, if $\mathcal{L}$ vanishes at one value $\lambda$ it vanishes everywhere, for any $C^2$ solution $(\Lambda, \Gamma)$. ($\mathcal{R}$ and $\mathcal{F}$ assumed in $C^0$.) Using that the real part of $\mathcal{L}$ is zero and that $\Gamma$ and $\Lambda$ have to fulfill the constraint equation (3.7), yields $\mathcal{L} = 0$ for a physical solution of (3.11). In terms of $\Gamma$, $\Lambda$ and their derivatives, $\sigma$ and $\theta$ can be written as

$$\theta = \frac{\dot{\Lambda}\Lambda^* - \dot{\Gamma}^*\Gamma}{|\Lambda|^2 - |\Gamma|^2} \quad , \quad \sigma = \frac{\dot{\Lambda}\Gamma^* - \Lambda\dot{\Gamma}^*}{|\Lambda|^2 - |\Gamma|^2} \quad , \tag{3.13}$$

provided the Jacobi map does not become singular; note that the reality of $\theta$ is equivalent to $\mathcal{L} = 0$. Therefore, one can obtain the series-expansions of $\theta$ and $\sigma$ by inserting the expansions of $\Gamma$ and $\Lambda$ which are derived from (3.11).

**The light beam at a focus**

At a focus, $\Lambda_c = 0 = \Gamma_c$; from our theorem we know that $|\dot{\Gamma}_c| \neq |\dot{\Lambda}_c|$ (otherwise the rank of $\dot{\mathcal{D}}_c$ would be smaller than two). We obtain from (3.11)

$$\Lambda(\epsilon) = \epsilon \dot{\Lambda}_c + \frac{\epsilon^3}{6}\left(\mathcal{R}_c \dot{\Lambda}_c - \mathcal{F}_c \dot{\Gamma}_c\right) + \mathcal{O}(\epsilon^4) \quad , \tag{3.14a}$$

$$\Gamma(\epsilon) = \epsilon \dot{\Gamma}_c + \frac{\epsilon^3}{6}\left(\mathcal{R}_c \dot{\Gamma}_c - \mathcal{F}_c^* \dot{\Lambda}_c\right) + \mathcal{O}(\epsilon^4) \quad , \tag{3.14b}$$

and thus the determinant of the Jacobian becomes

$$\det \mathcal{D}(\epsilon) = \epsilon^2 \left[1 + \frac{\epsilon^2}{3}\mathcal{R}_c\right] \det \dot{\mathcal{D}}_c + \mathcal{O}(\epsilon^5) \quad . \tag{3.15}$$

Since $\det \dot{\mathcal{D}}_c \neq 0$, the leading term of $\det \mathcal{D}(\epsilon)$ is of second order. The optical scalars become near the focus

$$\theta = \frac{1}{\epsilon}\left[1 + \frac{1}{3}\mathcal{R}_c \epsilon^2\right] + \mathcal{O}(\epsilon^2) \quad , \quad \sigma = \frac{1}{3}\mathcal{F}_c \epsilon + \mathcal{O}(\epsilon^2) \quad . \tag{3.16a\&b}$$

The function $w$ defined in (2.9) is equal to

$$w(\epsilon) = \text{sign}\left(\det \dot{\mathcal{D}}_\text{c}\right) |\epsilon| \sqrt{\left|\det \dot{\mathcal{D}}_\text{c}\right|} \left(1 + \mathcal{O}(\epsilon^2)\right) \quad , \tag{3.17a}$$

thus, it is continuous but not $C^1$ at the focus; $\dot{w}$ has a finite discontinuity. One obtains the expansions of $\Gamma$ and $\Lambda$ near the vertex by inserting the special values $\dot{\Gamma}_\text{v} = 0 = \mathcal{I}\text{m}\dot{\Lambda}_\text{v}$ and $\mathcal{R}e\dot{\Lambda}_\text{v} = 1$ into (3.14). As expected we obtain for $w$ at the vertex:

$$w(\epsilon) = \epsilon \left(1 + \mathcal{O}(\epsilon^2)\right) \quad , \quad \epsilon > 0 \quad . \tag{3.17b}$$

As already claimed, the optical scalars (3.16) have the same structure at a vertex and at a focus, since the expansions of the light beam around these points differ only by the opening angle $\det \dot{\mathcal{D}}_\text{v} = 1$ and $\det \dot{\mathcal{D}}_\text{c}$ which cancels in the numerator and denominator of $\theta$ and $\sigma$. Note that in lowest order ($\epsilon^{-1}$) the behaviour of $\theta$ and $\sigma$ at the vertex (focus) is expected: the infinitesimal neighborhood of such an event can be treated asymptotically as the flat Minkowski spacetime. In Minkowski spacetime, however, $\theta(\lambda) = \frac{1}{\lambda}$ and $\sigma(\lambda) \equiv 0$ holds for all $\lambda$, in particular at the vertex. The first order terms in $\theta$ and $\sigma$ demonstrate that the source of convergence $\mathcal{R}_0 \leq 0$ at the vertex (or focus) decreases the divergence of a beam, and that the source of shear produces a shear rate $\sigma$:

$$\dot{\sigma}(0) = \frac{1}{3}\mathcal{F}_0 \quad . \tag{3.18}$$

This implies: $\mathcal{F}_0 = 0 \iff \dot{\sigma}(0) = 0$, and with (2.2d), $\mathcal{F} \equiv 0 \iff \sigma \equiv 0$. Thus, a beam centred on $\gamma_0$ is shear free, if and only if the tangent vector of $\gamma_0$ is one of the at most 4 principal null directions of the conformal tensor, a rare, exceptional case. Thus generically $\sigma \neq 0$.

The fact that $\sigma = 0$ at the vertex implies that the coefficient of the rhs of the focusing equation (2.8) is continuous at the vertex; thus its solution $w$ is well defined at least from the observer to the first conjugate point.

**The light beam at a non-degenerate conjugate point**
At a non-degenerate conjugate point, the local expansion of the beam is determined by $\Gamma_\text{c} = \Lambda_\text{c} \neq 0$, $\dot{\Lambda}_\text{c}$ and $\dot{\Gamma}_\text{c}$. Since the constraint equation (3.7) has to be satisfied, there are only five free initial conditions: let $a$ and $b$ be the unique complex numbers which satisfy $\dot{\Lambda}_\text{c} = a\Lambda_\text{c}$ and $\dot{\Gamma}_\text{c} = b\Lambda_\text{c}$; then (3.7) yields $\mathcal{I}\text{m}\left[a + b\right] = 0, \Rightarrow \mathcal{R}e\left[a - b\right] = a - b^*$. The zero eigenvector of $\mathcal{D}_\text{c}$ is not a zero eigenvector of $\dot{\mathcal{D}}_\text{c}$ if and only if $\mathcal{R}e\left[a - b\right] \neq 0$; therefore $\mathcal{R}e\left[a\right] \neq \mathcal{R}e\left[b\right]$. With (3.11), the expansions of $\Lambda$ and $\Gamma$ near the conjugate point can be written as

$$\Lambda(\epsilon) = \left[1 + a\epsilon + \frac{1}{2}\epsilon^2 \left(\mathcal{R}_\text{c} - \mathcal{F}_\text{c}\right)\right]\Lambda_\text{c} + \mathcal{O}(\epsilon^3) \quad , \tag{3.19a}$$

$$\Gamma(\epsilon) = \left[1 + b\epsilon + \frac{1}{2}\epsilon^2 \left(\mathcal{R}_\text{c} - \mathcal{F}_\text{c}^*\right)\right]\Lambda_\text{c} + \mathcal{O}(\epsilon^3) \quad . \tag{3.19b}$$

The determinant of the Jacobian matrix is equal to

$$\det \mathcal{D}(\epsilon) = 2\mathcal{R}e\left[a - b\right]|\Lambda_\text{c}|^2 \epsilon + \left[|a|^2 - |b|^2\right]|\Lambda_\text{c}|^2 \epsilon^2 + \mathcal{O}(\epsilon^3) \quad ; \tag{3.20}$$

thus, the leading order of this expansion is equal to one. For the optical scalars, we obtain,

$$\theta(\epsilon) = \frac{1}{2\epsilon} \left[ 1 + \frac{|a|^2 - |b|^2}{2\mathcal{R}\mathrm{e}\,[a-b]} \epsilon + \mathcal{O}(\epsilon^2) \right] \quad , \tag{3.21a}$$

$$\sigma(\epsilon) = \frac{1}{2\epsilon} \left[ 1 - \frac{|a|^2 - |b|^2}{2\mathcal{R}\mathrm{e}\,[a-b]} \epsilon + \mathcal{O}(\epsilon^2) \right] \quad ; \tag{3.21b}$$

hence, the rate of shear is real (in the chosen coordinate frame, in which $\Lambda_\mathrm{c} = \Gamma_\mathrm{c}$) in zeroth order, it becomes imaginary in first order if and only if $\mathcal{F}_\mathrm{c}$ is a not real. The function $w$ is

$$w(\epsilon) = |\Lambda_\mathrm{c}| \operatorname{sign}(\mathcal{R}\mathrm{e}\,[a-b]\,\epsilon) \sqrt{|2\mathcal{R}\mathrm{e}\,[a-b]\,\epsilon|} \left(1 + \mathcal{O}(\epsilon^2)\right)$$

or with the abbreviation $\mathrm{d}_\mathrm{c} = \left[\frac{\mathrm{d}}{\mathrm{d}\lambda} \det \mathcal{D}\right]_{\lambda_\mathrm{c}}$:

$$w(\epsilon) = \operatorname{sign}(\mathrm{d}_\mathrm{c}\epsilon) \sqrt{|\mathrm{d}_\mathrm{c}\epsilon|} \left(1 + \mathcal{O}(\epsilon^2)\right) \quad .$$

Thus at a non-degenerate conjugate point $w$ is continuous, changes its sign, and has an infinite first derivative.

Now we summarize the results for the behaviour of the determinant of the Jacobi map and the optical scalars near conjugate points:
(1) at a non-degenerate conjugate point, $\det \mathcal{D} \propto \epsilon$, $\theta = 1/2\epsilon$, $\sigma \propto 1/2\epsilon$; in leading order, and
(2) at a focus, $\det \mathcal{D} \propto \epsilon^2$, $\theta = 1/\epsilon$, $\sigma = 0$.
With our knowledge of the behaviour of the shear rate $\sigma$ at a conjugate point, we now can prove that the focusing equation (2.8b) is integrable over the singularity at a conjugate point: In the worst case, that rhs of (2.8b) behaves like $|\sigma|^2 \sqrt{\det \mathcal{D}} \propto \epsilon^{-3/2}$; this yields $w(\epsilon) \propto \operatorname{sign}(\epsilon)\sqrt{|\epsilon|}$. Thus the solution is well defined, even for the case where there is a conjugate point between source and observer. The behaviour of the determinant of the Jacobian map at the two different types of conjugate points also varifies that the sign of $w$ from (2.8) changes only at a non-degenerate conjugate point, as was claimed in Sect.2. Our results also show that the points of $\gamma_0$ conjugate to the vertex form a discrete set.

# 4 The derivation of the gravitational lens equation from geometrical optics

So far, no approximation was used. To evaluate the propagation equation (2.4) in an inhomogeneous universe requires several approximation assumptions. These will be stated in this chapter, and used to rederive the basic relations of the standard gravitational lens theory formalism from general relativity.

**The Friedmann universe**
If one assumes that the universe is isotropic and homogeneous, then its metric is given by the Robertson-Walker-metric. The only non-vanishing components of the metric tensor then are $g_{tt} = c^2$, $g_{ii} = -R^2(t)\tilde{g}_{ii}$, with $\tilde{g}_{rr} = \frac{1}{1-kr^2}$, $\tilde{g}_{\theta\theta} = r^2$ and $\tilde{g}_{\phi\phi} = r^2(\sin\theta)^2$; the value of $k = 0, +1, -1$ determines whether the space is flat, spherical or hyperbolic; $t$ is the cosmic time. A fundamental observer with four velocity $U^\alpha(\lambda)$ at an event $\lambda$ on the central ray of a beam measures the frequency $\omega(\lambda) := ck^\alpha(\lambda)U_\alpha(\lambda) = -\omega_0 \tilde{k}^\alpha(\lambda)U_\alpha(\lambda) =:$

$\omega_0 (1 + z(\lambda))$; $k^\alpha$ is the wavevector of the central ray, $\omega_0$ is the frequency at the vertex of the beam and $z(\lambda)$ is (by definition) the (red)shift. In a Robertson-Walker-metric, the redshift is isotropic and is related to the scale factor of the metric by $R(z) = \frac{R_0}{1+z}$, where $R_0$ is the scale factor at the vertex of the beam ($z = 0$, $t = t_0$). The affine parameter-redshift differential equation is

$$\frac{\mathrm{d}z}{\mathrm{d}\lambda} = \frac{1}{c} \frac{\frac{\mathrm{d}}{\mathrm{d}t}R(t)}{R(t)} [1 + z(\lambda)]^2 = \frac{1}{c} \frac{\frac{\mathrm{d}}{\mathrm{d}t}R(t)}{R(t_0)} [1 + z(\lambda)]^3 \quad . \tag{4.1}$$

Note that this yields a proper distance-affine parameter relation at redshift $z$ of

$$\mathrm{d}D_{\mathrm{proper}} = (1 + z)\mathrm{d}\lambda \quad , \tag{4.2}$$

which is consistent with our convention that the affine parameter equals the proper length at the vertex at $\lambda = 0 = z$. For a *Friedmann universe* with zero cosmological constant and an energy momentum tensor of a matter-dominated ideal fluid, $p \ll \rho c^2$, equation (4.1) can be solved by inserting the Friedmann equation for $\frac{\dot{R}(t)}{R(t)}$:

$$\lambda(z) = \frac{c}{H_0} \int_0^z \frac{\mathrm{d}z'}{(1 + z')^3 \sqrt{\Omega z' + 1}} \quad ; \tag{4.3}$$

$H_0$ is the Hubble parameter $\mathrm{d}(\ln R)/\mathrm{d}t$ at the observation event $t_0$.

**Parallel transport in a Robertson-Walker spacetime**
To calculate the source of shear defined in (2.2g), we need the screen vectors $E_i^\alpha$, $i = 1, 2$, and $\tilde{k}^\alpha$ along the central ray. We choose the center of the spatial coordinate system $(r, \theta, \phi)$ at the observer, and the central ray $\gamma_0$ connecting source and observer in the direction of $\theta = \frac{\pi}{2}$. Consider the dimensionless function

$$\Upsilon(t, r) := \int_{t_0}^t \frac{c\mathrm{d}\tau}{R(\tau)} + \int_0^r \frac{\mathrm{d}x}{\sqrt{1 - kx^2}} \quad .$$

It solves the eikonal equation; the hypersurface $\Upsilon(t, r) = \Upsilon(t', 0)$ defines the past null cone of $(t', 0)$. Therefore, the phase functions converging on the world line $r = 0$ are all given by $S(t, r) = f(\Upsilon(t, r))$, where $f$ depends on the phase $S(t, 0)$. The vector $\tilde{k}_\alpha$ (which is on $\mathcal{C}_0^-$) has to be a constant multiple of $\Upsilon_{,\alpha} = (R^{-1}(t), 1/\sqrt{1 - kr^2}, 0, 0)$;[5] since $\tilde{k}_0 = -1$ at the vertex, we obtain

$$\tilde{k}_\alpha(z) = -(1 + z) \left(1, \frac{R}{\sqrt{1 - kr^2}}, 0, 0\right)$$

and thus

$$\tilde{k}^\alpha(z) = (1 + z) \left[-1, \frac{1}{\sqrt{-g_{rr}(z)}}, 0, 0\right] \quad . \tag{4.4}$$

The spacelike screen vectors $E_1^\alpha$ and $E_2^\alpha$ adapted to $\tilde{k}^\alpha$ can be chosen at the observer proportional to $[0, 0, 1, 0]$ and $[0, 0, 0, 1]$. For general $z$ we then obtain

---
[5] The components of a four vector $x^\alpha$ are $x^0 = ct, r, \theta, \phi$.

$$E_1^\alpha(z) = \frac{1}{\sqrt{-g_{\theta\theta}(z)}}[0,0,1,0] \quad , \quad E_2^\alpha(z) = \frac{1}{\sqrt{-g_{\phi\phi}(z)}}[0,0,0,1] \quad . \tag{4.5}$$

The components of the vectors $E_i^\alpha(0)$ become singular at the observer at $z=0$. This is due to the choice of the coordinate system; the vectors themselves and their inner products are regular.

**The on-average Friedmann universe**
Of course, a homogeneous universe is not realistic. A better model must take into account that only a fraction $0 \leq \tilde{\alpha} \leq 1$ of the matter is distributed homogeneously, whereas the rest is concentrated in clumps. Imagine a model universe that is inhomogeneous on small scales and homogeneous on large scales (some 100 Mpc's) such that this clumping of matter does not affect "global" (or large scale) functions like $R(t)$, $R(z)$, $\lambda(z)$ and the parallelly transported fields $\tilde{k}^\alpha(z)$, $E_{(i)}^\alpha(z)$. This means that, on average, this universe behaves like a Friedmann universe with density $\rho_F$ which has the same total matter content as the actual universe. Thus, such a model is called an *on-average Friedmann universe* (see, e.g. Zeldovich 1964, Dyer & Roeder 1973).

This picture of the matter distribution in our universe is a realistic one if one is interested in the light deflection caused by 'strong', isolated matter inhomogeneities, such as galaxies and clusters of galaxies, the deflectors which produce multiply-imaged QSOs, radio rings, and luminous arcs. In these situations, it seems to be a fair approximation to consider the light beams between us and the deflector, and between the deflector and the source to be nearly unperturbed by matter inhomogeneities; if there is more than one deflector along the line-of-sight, this can be accounted for in the present prescription. An alternative view of the matter distribution in the universe is provided by considering larger scales, on which the density inhomogeneities are linear or quasi-linear. Then it is more realistic to model the matter distribution as a field $\delta\rho$ which is superposed on the Friedmann density $\rho_F$, such that $\langle\delta\rho\rangle = 0$, and the average is taken on spatial scales which are small compared to the Hubble length, but larger than the largest scale on which the density fluctuations $\delta\rho$ still have appreciable power (see, e.g., Gunn 1967, Blandford et al. 1991, Kaiser 1992 for studies of light propagation in such a weakly inhomogeneous universe). In the following we adopt the first view, that of a clumpy universe; we note, however, that most of our results derived below also apply for the weakly inhomogeneous universe. In particular, the (multiple deflection) gravitational lens equation can also be used in the latter case, if the universe is 'sliced' into redshift bins and the matter inhomogeneities are projected onto 'lens planes' in the bins, since the multiple deflection gravitational lens equation can be considered just as a discretization of the exact propagation equation (2.4). The only modification that has to be applied in the case of a weakly inhomogeneous universe is that $\mathcal{R}_{\rm cl}$ no longer is nonpositive, and the projected surface mass density $\Sigma$ in each lens plane can attain positive and negative values. Furthermore, since the magnification, defined in Sect. 5 below, is defined relative to the Friedmann-Lemaître universe, the mean magnification relative to that must be unity (see the discussion in Sect. 4.5.1 of SEF), and the focusing theorem of Sect. (5.12) no longer holds, since $\mathcal{R}_{\rm cl}$ can have either sign.

## 4.1 The sources of shear and convergence for weak, isolated inhomogeneities

**Weak, isolated inhomogeneities**

We assume that inhomogeneities like galaxies or clusters of galaxies are isolated from each other such that in each domain containing an inhomogeneity, small compared to the Hubble distance, the metric can be approximated by a post-Minkowskian line element

$$ds^2 = \left(1 + 2\frac{\Phi}{c^2}\right) c^2 dt^2 - \left(1 - 2\frac{\Phi}{c^2}\right) d\mathbf{x}^2 \quad .$$

The relative velocities of its mass distribution are small, $v \ll c$, and its Newtonian gravitational potential $\Phi$ is weak, $\Phi \ll c^2$. If the density outside such regions is $\tilde{\alpha}\rho_F$ and we write for the density inside a clump $\tilde{\alpha}\rho_F + \rho_{cl}$, such that $\rho_{cl}$ is localized in the region, Poissons's equation $\Delta_3 \Phi = 4\pi G \rho_{cl}$ holds within the region.[6] The metric does not change appreciably on the time scale light needs to propagate through the inhomogeneity. We therefore call such inhomogeneities *quasistatic, weak inhomogeneities*.

**The source of convergence**

First we consider the source of convergence $\mathcal{R}$, defined in (2.2f). Inserting the field equations with an energy-momentum tensor of an ideal fluid yields:

$$\mathcal{R} = -\frac{4\pi G}{c^2} \tilde{\rho} \tilde{U}_\alpha \tilde{U}_\beta \tilde{K}^\alpha \tilde{K}^\beta \quad . \tag{4.6}$$

In this equation, $\tilde{U}^\alpha$ is the four velocity of the ideal fluid, which deviates from the velocity in a pure Friedmann universe $U^\alpha$ by the peculiar velocity $U^\alpha_{pec}$, $\tilde{U}^\alpha = U^\alpha + U^\alpha_{pec}$, and $\tilde{K}^\alpha$ is the wave vector of the central ray of the beam considered, which deviates from the wavevector $\tilde{k}^\alpha$ in a Friedmann universe due to deflection in the inhomogeneity by a vector $\delta \tilde{k}^\alpha$, $\tilde{K}^\alpha = \tilde{k}^\alpha + \delta \tilde{k}^\alpha$. The matter density $\tilde{\rho} = \rho_{bg} + \rho_{cl}$ is given as a sum of the reduced background density in the on-average Friedmann universe, $\rho_{bg} = \tilde{\alpha}\rho_F$, and the matter density of the clump $\rho_{cl}$. If we use that peculiar velocities of inhomogeneities (e.g., galaxies) are small, $v_{pec} \lesssim 10^{-3} c$, and that their gravitational fields are also small, $\delta z_g \approx \frac{2\Phi}{c^2} \ll 1$, we can neglect the contributions from $U^\alpha_{pec}$ and $\delta k^\alpha$ and obtain from (4.6) that in lowest order, with $\mathcal{R} = \mathcal{R}_{bg} + \mathcal{R}_{cl}$, the contribution of the clumps is given by

$$\mathcal{R}_{cl} \approx -\frac{4\pi G}{c^2} \rho_{cl} U_\alpha U_\beta \tilde{k}^\alpha \tilde{k}^\beta \quad . \tag{4.7}$$

Consider an inhomogeneity along a ray $\gamma_0$ localized in the affine parameter interval $[\lambda_{min}, \lambda_{min} + \Delta\lambda]$ which is small compared to its distance to us: $\Delta\lambda \ll \lambda_{min}$; let $z_d$ be an element of the corresponding redshift interval $[z_{min}, z_{min} + \Delta z]$. Since the inhomogeneity must not change significantly during the time the light beam traverses it, we can calculate (4.7) for one instant of time, $t(z_d)$. The line element in the asymptotically flat neighborhood $\mathcal{U}$ of $\lambda(z_d)$ is $ds^2 = \left(1 + 2\frac{\Phi}{c^2}\right)(c\,dt)^2 - \left(1 - 2\frac{\Phi}{c^2}\right)(d\boldsymbol{\zeta}^2)$, with $t = R_d \eta$, $R_d^2 d\sigma_k^2 \approx (d\boldsymbol{\zeta})^2$ and $R(z_d) = R_d$; $(t, \boldsymbol{\zeta})$ denote Post-Minkowski-coordinates centered on $\lambda(z_d)$ and oriented such that $\zeta_3$ is parallel to the spatial direction of $\gamma_0$ there. We calculate $\mathcal{R}$ and $\mathcal{F}$ not only on the central ray $\gamma_0$ of the beam considered, but for all spatial positions $\boldsymbol{\zeta}$ in $\mathcal{U}$. This yields $\mathcal{R}$ and $\mathcal{F}$ for all rays traversing $\mathcal{U}$, where the spatial paths of the rays are parametrized by $\boldsymbol{\zeta}(\lambda)$; note that the rays in $\mathcal{U}$ do not have to be infinitesimally near to $\gamma_0$ in the sense of (2.3). The source of convergence on a ray in $\mathcal{U}$ is the sum of

$$\mathcal{R}_{\mathrm{bg}}(\lambda) = -\frac{4\pi G}{c^2}\rho_{\mathrm{bg}}(z)\left[1+z\right]^2 \quad , \quad \mathcal{R}_{\mathrm{cl}}(\lambda) = -\frac{4\pi G}{c^2}\rho_{\mathrm{cl}}(\boldsymbol{\zeta}(\lambda))\left[1+z\right]^2 \quad , \qquad (4.8a)$$

where we have written $z$ instead of $z(\lambda)$. If one uses the Poisson equation $\Delta_3 \Phi = 4\pi G \rho_{\mathrm{cl}}$ for the quasistationary Newtonian gravitational potential $\Phi(t_{\mathrm{d}}, \boldsymbol{\zeta}) \approx \Phi(\boldsymbol{\zeta})$ of the inhomogeneity, this yields for $\mathcal{R}_{\mathrm{cl}}$:

$$\mathcal{R}_{\mathrm{cl}}(\lambda) = -\frac{(1+z)^2}{c^2}\Delta_3\Phi(\boldsymbol{\zeta}(\lambda)) \quad . \qquad (4.8b)$$

---

[6] Concerning the difficult problem of constructing approximate solutions to Einstein's equations containing quasi-static, weak inhomogeneities seperated by 'empty regions' and being Friedmannian on a large scale, see Futamase & Sasaki 1989, Jacobs et al 1993; see also Kasai 1993.

Up to now we have considered weak inhomogeneities which are small in size compared to their distance to us. Now we will restrict ourselves to those which are sufficiently thin, such that one can replace the wavevector and the vectors $E^\alpha$ in (2.2g) by (4.4) and (4.5) evaluated at the redshift of the clump. (That is, for the calculation of the source term for the evolution of the light beam, one can neglect the deflection relative to the unperturbed light beam). Thus we approximate

$$\boldsymbol{\zeta}(\lambda) \approx (\zeta_1(\lambda_{\mathrm{d}}), \zeta_2(\lambda_{\mathrm{d}}), \zeta_3(\lambda)) \qquad (4.9)$$

for rays which are roughly parallel to $\gamma_0$ at $\lambda(z_{\mathrm{d}})$; the deviation of rays from the parallel direction must be small, as well as the typical deflection angle caused by an inhomogeneity, otherwise the approximation (4.9) would break down.[7] With our choice of the coordinate system, $\zeta_1$ and $\zeta_2$ are orthogonal coordinates on the screen defined in Sect. 2. Therefore we write $(\zeta_1(\lambda_{\mathrm{d}}), \zeta_2(\lambda_{\mathrm{d}})) = \boldsymbol{\xi}$ in the following; $\boldsymbol{\xi}$ is a parameter to label rays. Equations (4.8) hold for an infinitesimal beam with central ray $\gamma_0$ ($\boldsymbol{\xi} = \mathbf{0}$), and for any other ray which is in $\mathcal{U}$ and roughly parallel to $\gamma_0$ at $\lambda_{\mathrm{d}}$.

The approximation (4.9) is equivalent to one on which gravitational lens theory is based: there, the source term for the light bending along the deflected light ray is approximated locally by that evaluated along the path of the unperturbed ray.

**The source of shear**
Outside the matter inhomogeneities, where $\rho_{\mathrm{bg}} = \tilde{\alpha}\rho_F$, we neglect the source of shear due to clumps; i.e., we neglect the long-range gravitational action of the weak inhomogeneities, and put $\mathcal{F} = 0$. At the inhomogeneity we evaluate (2.2g) in post-Minkowskian coordinates , hence we have to transform the coordinates from $(x^0, r, \theta, \phi)$ to $(x^0, \zeta_1, \zeta_2, \zeta_3)$. Note that we have chosen the $\zeta_3$-direction of the new coordinate system parallel to the spatial direction of the central ray. Since the normalization of all vectors stays invariant under the transformation of the coordinate system and since the norm in the local Minkowski-system is built with $\eta = \mathrm{diag}\,(1, -1, -1, -1)$, we have to replace the metric tensor $g$ by $\eta$ in (4.4) and (4.5) and we obtain

$$\tilde{k}^\alpha(z_{\mathrm{d}}) = -(1+z_{\mathrm{d}})\left[1,0,0,-1\right] \quad , \quad E_1^\alpha(z_{\mathrm{d}}) = [0,1,0,0] \quad , \quad E_2^\alpha(z_{\mathrm{d}}) = [0,0,1,0] \quad . \qquad (4.10)$$

---

[7] In astrophysically relevant situations, the beams under consideration have an opening angle of $\sim 1$ arcminute $\sim 3 \times 10^{-4}$ for galaxy clusters, and of $\sim 10$ arcseconds $\sim 5 \times 10^{-5}$ for lensing by individual galaxies; the corresponding typical deflection angles are of the same order or smaller.

The Riemann-tensor in the post-Minkowskian-approximation in $ct$ and $\zeta$ coordinates is equal to

$$R_{\alpha\beta\gamma\delta} = -\frac{1}{c^2}\left\{\delta_{\alpha\gamma}\Phi_{,\beta\delta} - \delta_{\beta\gamma}\Phi_{,\alpha\delta} - \delta_{\alpha\delta}\Phi_{,\beta\gamma} + \delta_{\beta\delta}\Phi_{,\alpha\gamma}\right\} \quad . \tag{4.11}$$

Thus, (4.10) and (4.11) yield that there are only contributions to the source of shear in (2.2g) if $\alpha,\gamma \in \{1,2\}$ *and* $\beta,\delta \in \{0,3\}$; hence the summation contains only 16 nonvanishing contributions. Using the quasistationarity of the metric, $\Phi_{,0} \ll \Phi_{,i}$, yields that in lowest order of $\frac{v}{c}$, only the following eight components of the curvature tensor contribute to (2.2g):

$$R_{1010} = -\frac{1}{c^2}\Phi_{,11} \quad , \quad R_{1020} = R_{2010} = -\frac{1}{c^2}\Phi_{,12} \quad , \quad R_{1313} = -\frac{1}{c^2}(\Phi_{,33} + \Phi_{,11}) \quad , \tag{4.12a}$$

$$R_{2020} = -\frac{1}{c^2}\Phi_{,22} \quad , \quad R_{1323} = R_{2313} = -\frac{1}{c^2}\Phi_{,12} \quad , \quad R_{2323} = -\frac{1}{c^2}(\Phi_{,33} + \Phi_{,22}) \quad . \tag{4.12b}$$

Inserting $\Phi_{,12} = \Phi_{,21}$, (4.12) and (4.10) in (2.2g) and using (4.9) yields

$$\mathcal{F}_{\rm cl}(\boldsymbol{\xi};\lambda) = \frac{1}{c^2}(1+z)^2 \left\{\Phi_{,11} - \Phi_{,22} - 2{\rm i}\Phi_{,21}\right\}(\boldsymbol{\xi};\zeta_3(\lambda)) \quad . \tag{4.13}$$

Therefore we obtain with (4.13), (4.8a) and (4.8b) that the optical tidal matrix along a *family* of rays traversing an asymptotically flat neighborhood of an event $\lambda_{\rm d}$ localized in a weak geometrically-thin clump in an on-average Friedmann universe, such that their spatial directions are roughly parallel to the $\zeta_3$-direction at $\lambda_{\rm d}$, is $\mathcal{T}(\boldsymbol{\xi};\lambda) = \mathcal{T}_{\rm bg}(z) + \mathcal{T}_{\rm cl}(\lambda)$ with $\mathcal{T}_{\rm bg}(z) = \mathcal{R}_{\rm bg}(z)\mathcal{I}$ and

$$(\mathcal{T}_{\rm cl})_{ik}(\boldsymbol{\xi};\lambda) = -\frac{(1+z)^2}{c^2}\left[2(\Phi_{,ik}) + (\delta_{ik}\Phi_{,33})\right](\boldsymbol{\xi};\zeta_3(\lambda)) \quad , \quad i,k \in \{1,2\} \quad . \tag{4.14}$$

Thus, the optical tidal matrix is simply related to the ordinary tidal matrix, i.e., the matrix of the second derivatives of the Newtonian potential. In these equations, $z = z(\lambda)$, and $\boldsymbol{\xi}$ is the screen position of the ray considered at $\lambda_{\rm d}$ relative to one chosen ray $\gamma_0$ of the family; $\zeta_3$ is the direction in the post-Minkowski coordinate system parallel to the rays at $\lambda_{\rm d}$, hence with (4.2)

$$\mathrm{d}\zeta_3 = (1+z)\mathrm{d}\lambda \quad . \tag{4.15}$$

If one evaluates the mapping of an infinitesimally thin beam (i.e., one needs the value of (4.14) on one ray $\gamma_0$ only), one puts $\boldsymbol{\xi} = \mathbf{0}$ in (4.14).

### 4.2 The thin lens approximation

One of the simplifying assumptions underlying lens theory is that the inhomogeneities are geometrically thin. Thus one approximates the inhomogeneities by two-dimensional *surface mass densities* $\Sigma$. Let one of the distributions be situated on the 'plane' $\zeta_3 = 0$,

$$\rho_{\rm cl}(\boldsymbol{\xi},\zeta_3) \approx \delta(\zeta_3)\Sigma(\boldsymbol{\xi}) \quad , \tag{4.16}$$

where

$$\Sigma(\boldsymbol{\xi}) := \int_{-\infty}^{+\infty} \mathrm{d}\zeta_3\, \rho_{\rm cl}(\boldsymbol{\xi},\zeta_3) \tag{4.17}$$

The Newtonian potential of this distribution is

$$\Phi(\boldsymbol{\xi}, \zeta_3) = -G \int \frac{\Sigma(\boldsymbol{\eta}) \mathrm{d}^2 \eta}{\sqrt{(\boldsymbol{\xi} - \boldsymbol{\eta})^2 + \zeta_3^2}} \quad . \tag{4.18}$$

The derivatives $\Phi_{,ik}$, $\Phi_{,33}$ which occur in the tidal matrix (4.14) decrease like the inverse third power of the distance from the plane of the mass distribution. It is, therefore, reasonable to approximate the optical tidal matrix for a clump of matter as a delta-distributional source term in $\lambda$:

$$\mathcal{T}_{\mathrm{cl}}^{\mathrm{p}}(\boldsymbol{\xi}; \lambda) := \delta(\lambda - \lambda_{\mathrm{d}}) \int_{-\infty}^{+\infty} \mathcal{T}_{\mathrm{cl}}(\boldsymbol{\xi}; \zeta_3(\lambda')) \, \mathrm{d}\lambda' \quad . \tag{4.19}$$

**The deflection potential $\tilde{\Psi}$**

The *deflection potential* $\tilde{\Psi}(\boldsymbol{\xi})$ of an inhomogeneity is defined as usual by

$$\tilde{\Psi}(\boldsymbol{\xi}) = \frac{4G}{c^2} \int \mathrm{d}^2 \xi' \, \Sigma(\boldsymbol{\xi}') \ln\left\{ \frac{|\boldsymbol{\xi} - \boldsymbol{\xi}'|}{D_{\mathrm{d}}} \right\} \quad ; \tag{4.20}$$

(see SEF, Sects. 4.3 & 5.1). In the deflection potential, the denominator in the argument of the logarithm is an arbitrary length, to make this argument dimensionless; we have chosen it equal to the so-called empty cone angular diameter distance $D_{\mathrm{d}} := D(z_{\mathrm{d}})$ from the observer to the redshift $z_{\mathrm{d}}$. Under a change of this length scale, the value of (4.20) changes only by an unimportant additive constant. It is straightforward to see that $\tilde{\Psi}$ and $\Sigma$ are related to each other by the Poisson equation for the surface mass density

$$\Delta_{(2)} \tilde{\Psi}(\boldsymbol{\xi}) = \frac{8\pi G}{c^2} \Sigma(\boldsymbol{\xi}) \quad , \tag{4.21}$$

where $\Delta_2$ is the two-dimensional Laplace operator.

We now show that the approximate tidal matrix of eq. (4.19) can be expressed in terms of the second derivatives of the deflection potential rather than in terms of the $\Phi$-derivatives. In fact, using eqs. (4.18), (4.20) and (4.15) one verifies by a straightforward calculation that, for $i, k \in \{1.2\}$,

$$\int_{-\infty}^{\infty} \mathrm{d}\zeta_3 \, \Phi_{,ik}(\boldsymbol{\xi}, \zeta_3) = \frac{c^2}{2} \tilde{\Psi}_{,ik}(\boldsymbol{\xi}) \quad , \tag{4.22}$$

$$\int_{-\infty}^{\infty} \mathrm{d}\zeta_3 \, \Phi_{,33}(\boldsymbol{\xi}, \zeta_3) = 0 \quad . \tag{4.23}$$

Therefore, eq. (4.18) leads to

$$\mathcal{T}_{\mathrm{cl}}^{\mathrm{p}}(\boldsymbol{\xi}; \lambda) = -(1+z_{\mathrm{d}}) \, \delta(\lambda - \lambda_{\mathrm{d}}) \begin{bmatrix} \tilde{\Psi}_{,11}(\boldsymbol{\xi}) & \tilde{\Psi}_{,12}(\boldsymbol{\xi}) \\ \tilde{\Psi}_{,21}(\boldsymbol{\xi}) & \tilde{\Psi}_{,22}(\boldsymbol{\xi}) \end{bmatrix} = -(1+z_{\mathrm{d}}) \, \delta(\lambda - \lambda_{\mathrm{d}}) \, \tilde{U}(\boldsymbol{\xi}) \quad . \tag{4.24}$$

In the last step, we have defined the *deflection matrix* $\tilde{U}(\boldsymbol{\xi})$ as the Hesse-matrix of the deflection potential $\tilde{U}(\boldsymbol{\xi}) =: \mathcal{H}\left[\tilde{\Psi}\right](\boldsymbol{\xi})$.

We can generalize the result (4.24) to the case of several inhomogeneities, i.e., for the following case, which also is the "standard situation" in gravitational lens theory: given an observer at redshift zero in an on-average Friedmann universe, a source at redshift $z_s =: z_{N+1}$ and an arbitrary number $N$ of geometrically-thin, weak inhomogeneities between source and observer, situated at $\lambda_1,..,\lambda_N$ with corresponding redshifts of $z_1,..,z_N$. Then, if we indicate the two-dimensional screen positions of a ray (relative to one ray $\gamma_0$ of the family) in the inhomogeneities with $\boldsymbol{\xi}_j$ and the deflection matrices at those positions as $\tilde{U}_j(\boldsymbol{\xi}_j)$, the optical tidal matrix is equal to

$$\mathcal{T}^{\mathrm{p}}(\boldsymbol{\xi}_1,...,\boldsymbol{\xi}_N;\lambda) = \mathcal{R}_{\mathrm{bg}}(\lambda)\mathcal{I} - \sum_{i=1}^{N}(1+z_i)\,\tilde{U}_i(\boldsymbol{\xi}_i)\,\delta(\lambda-\lambda_i) \quad ; \tag{4.25a}$$

the different rays considered must be roughly parallel to each other before the first inhomogeneity, then, the same holds at every following inhomogeneity provided the deflection angles are small. Again, considering only one infinitesimal beam with central ray $\gamma_0$, one has to consider

$$\mathcal{T}^{\mathrm{p}}(\lambda) = \mathcal{R}_{\mathrm{bg}}(\lambda)\mathcal{I} - \sum_{i=1}^{N}(1+z_i)\,\tilde{U}_i(\mathbf{0})\,\delta(\lambda-\lambda_i) \quad . \tag{4.25b}$$

### 4.3 The recurrence relation for the mapping of the light beam

The equations (4.25) result from well-defined assumptions and approximations. Hence we can solve the differential equation (2.4a) with (4.25) as source term. We again consider not only a single beam, but a family of beams with (nearly) parallel central rays, and label a beam by the screen position $\boldsymbol{\xi}_n$ of its central ray relative to one reference ray $\gamma_0$. Defining $\dot{\mathcal{D}}_n^{+}(\boldsymbol{\xi}_n) := \lim_{\lambda \searrow \lambda_n} \dot{\mathcal{D}}(\boldsymbol{\xi}_n;\lambda)$ and $\dot{\mathcal{D}}_n^{-}(\boldsymbol{\xi}_n) := \lim_{\lambda \nearrow \lambda_n} \dot{\mathcal{D}}(\boldsymbol{\xi}_n;\lambda)$, this yields:

$$\dot{\mathcal{D}}_n^{+}(\boldsymbol{\xi}_n) - \dot{\mathcal{D}}_n^{-}(\boldsymbol{\xi}_n) = -(1+z_n)\tilde{U}_n(\boldsymbol{\xi}_n)\mathcal{D}_n(\boldsymbol{\xi}_n) \quad ; \tag{4.26}$$

thus the Jacobi matrix, but not its derivative is continuous at an inhomogeneity in lens approximation. On the lhs of (4.26) we want to express the derivatives of the Jacobi matrices as functions of the values of the Jacobi matrices at redshifts $z_{n-1}$, $z_n$ and $z_{n+1}$. In order to do this, we first have to determine the evolution of an infinitesimal light beam outside clumps.

**The evolution of a beam outside clumps, Dyer-Roeder differential equation**
We now investigate the evolution of a beam outside clumps, which we call empty beam or *empty cone* in the following. Since outside of clumps the source of shear vanishes, the differential equation (2.4a) simplifies with the first of (4.8a) to

$$\ddot{\mathcal{D}}(\lambda) = \mathcal{R}_{\mathrm{bg}}(\lambda)\mathcal{D}(\lambda) = -\frac{4\pi G}{c^2}\rho_{\mathrm{bg}}(z)\,[1+z]^2\,\mathcal{D}(\lambda) \quad .$$

If we insert the evolution of the density with redshift, $\rho_{\mathrm{bg}}(z) = \tilde{\alpha}\rho_0(1+z)^3$, the definition of the density parameter $\Omega = \frac{\rho_0}{\rho_{\mathrm{crit}}}$ with $\rho_{\mathrm{crit}} = \frac{3H_0^2}{8\pi G}$, we find that each component of $\mathcal{D}$ fulfills the differential equation

$$\frac{\mathrm{d}^2}{\mathrm{d}\lambda^2}B(\lambda) = -\frac{3}{2}\left(\frac{c}{H_0}\right)^{-2}\tilde{\alpha}\Omega\left[1+z(\lambda)\right]^5 B(\lambda) \quad .$$

Using the affine parameter-redshift relation (4.3), this finally transforms to the *Dyer-Roeder differential equation* (Dyer & Roeder 1973)

$$(\Omega z + 1)(1+z)\frac{\mathrm{d}^2 B(z)}{\mathrm{d}z^2} + \left[\frac{7}{2}\Omega z + \frac{\Omega}{2} + 3\right]\frac{\mathrm{d}B(z)}{\mathrm{d}z} + \frac{3}{2}\tilde{\alpha}\Omega B(z) = 0 \quad . \tag{4.27}$$

This second order differential equation has two linearly independent solutions; two solutions $B_1$ and $B_2$ are independent if and only if the Wronskian $W(z) := \dot{B}_1 B_2 - \dot{B}_2 B_1(z)$ is different from zero at one value of $z$ (and thus for every $z$). The first and second terms of equation (4.27) describe the evolution of a light beam due to the expansion of the universe, therefore $\Omega$ appears; the third term describes the convergence of a light beam due to the local homogeneous matter density $\tilde{\alpha}\rho_\mathrm{F}$ in the empty cone (no clumps); for this reason, a term $\tilde{\alpha}\Omega$ occurs. Consider a solution $D(z_i, z)$ of (4.27) which is zero at redshift $z_i$ and whose derivative with respect to redshift obeys the local Hubble law, or equivalently, the infinitesimal quantity $\frac{\mathrm{d}D}{\mathrm{d}z}\mathrm{d}z$ equals the infinitesimal proper length $\mathrm{d}D_\mathrm{proper}(z_i)$ at redshift $z_i$. Then $D(z_1, z_2)$ is the empty cone angular diameter distance from redshift $z_1$ to $z_2$; it can be described by a function $r(z_i, z)$, solving (4.27) with boundary conditions

$$\frac{\mathrm{d}}{\mathrm{d}z}r(z_i, z)|_{z=z_i} = \frac{1}{(1+z_i)^2\sqrt{\Omega z_i + 1}} \quad , \quad r(z_i, z)|_{z=z_i} = 0 \quad , \tag{4.28}$$

in the following form:

$$D(z_1, z_2) = \frac{c}{H_o}|r(z_1, z_2)| \quad . \tag{4.29}$$

The general solution of this initial value problem is provided in Seitz & Schneider (1994). If there is no inhomogeneity in the beam between its vertex and redshift $z$, the Jacobi matrix $\mathcal{D}(z)$ is given by $\mathcal{D}(z) = \frac{c}{H_0}r(0,z)\mathcal{I}$; in particular, at the first inhomogeneity at $z_1$, $\mathcal{D}(z_1) = D(0, z_1)\mathcal{I}$. To describe the solution of eq. (2.4a) between the $(n-1)$-th and $n$-th and between the $n$-th and $(n+1)$-th inhomogeneity, we put:

$$\mathcal{D}(z) = X_1 B_1(z) + Y B_2(z) \quad , \quad z \in [z_{n-1}, z_n] \quad , \tag{4.30}$$

$$\mathcal{D}(z) = X_2 B_1(z) + Z B_2(z) \quad , \quad z \in [z_n, z_{n+1}] \quad . \tag{4.31}$$

Here, $B_1$ and $B_2$ are linearly independent solutions of the Dyer-Roeder differential equation; we choose them as $B_1(z) := D(0, z) =: D(z)$ and $B_2(z) = D(z_n, z)$. $X_1$, $X_2$, $Y$ and $Z$ are real $2\times 2$-matrices, determined by the boundary conditions. Evaluating (4.30) and (4.31) at $z_n$ immediately yields $X := X_1 = X_2 = \frac{1}{D_n}\mathcal{D}_n =: A_n$ Then, we calculate the derivatives of (4.30) and (4.31) with respect to $\lambda$, evaluate these at $\lambda_n$ and obtain with (4.3) and (4.28) the difference:

$$\dot{\mathcal{D}}_n^+ - \dot{\mathcal{D}}_n^- = Z\frac{\mathrm{d}}{\mathrm{d}\lambda}D(z_n, z)|_{z\searrow z_n} - Y\frac{\mathrm{d}}{\mathrm{d}\lambda}D(z_n, z)|_{z\nearrow z_n} = (1+z_n)[Z+Y] \quad . \tag{4.32}$$

The matrices $Y$ and $Z$ can be calculated by evaluating (4.30) and (4.31) at $z_{n-1}$ and $z_{n+1}$, respectively. With the abbreviations $D(z_i, z_j) =: D_{ij}$ and $D(z_i) =: D_i$ this yields:

$$Y = \frac{1}{D_{n,n-1}}\left\{\mathcal{D}_{n-1} - \frac{D_{n-1}}{D_n}\mathcal{D}_n\right\} \quad , \quad Z = \frac{1}{D_{n,n+1}}\left\{\mathcal{D}_{n+1} - \frac{D_{n+1}}{D_n}\mathcal{D}_n\right\} \quad . \quad (4.33)$$

We insert (4.33) and (4.32) into (4.26), use the Etherington (1933) reciprocity relation

$$\frac{D(z_1,z)_{|z=z_2}}{1+z_1} = \frac{D(z_2,z)_{|z=z_1}}{1+z_2}$$

and obtain for $\mathcal{D}_{n+1}$:

$$\mathcal{D}_{n+1} = -D_{n,n+1}\tilde{U}_n\mathcal{D}_n - \frac{1+z_{n-1}}{1+z_n}\frac{D_{n,n+1}}{D_{n-1,n}}\mathcal{D}_{n-1} + \\ + \left\{\frac{(1+z_{n-1})}{1+z_n}\frac{D_{n,n+1}D_{n-1}}{D_{n-1,n}D_n} + \frac{D_{n+1}}{D_n}\right\}\mathcal{D}_n \quad . \quad (4.34)$$

However, this relation is equivalent to the recurrence relation for the Jacobi matrices in lens theory. This becomes clear, if one rewrites this equation, as common in lens theory, in dimensionless form. One has to insert the dimensionless deflection matrix $U(\mathbf{x})$ related to $\tilde{U}(\boldsymbol{\xi})$ via

$$U_j(\mathbf{x}_j) = \frac{D_{j,N+1}D_j}{D_{N+1}}\tilde{U}_j(\boldsymbol{\xi}_j) \quad , \quad \boldsymbol{\xi}_j =: \mathbf{x}_j D_j \quad ,$$

and the definitions of the dimensionless Jacobi matrices $A_i(\mathbf{x}_i) := \frac{1}{D_i}\mathcal{D}_i(D_i\mathbf{x}_i)$. Defining the geometrical quantities

$$\vartheta_i := \frac{(1+z_i)}{c}\frac{D_iD_{i+1}}{D_{i,i+1}} \quad , \quad 0 \leq i \leq N \quad , \quad v_i := -\sqrt{\frac{\vartheta_{i-1}}{\vartheta_i}} \quad , \quad 1 \leq i \leq N$$

and

$$\beta_{ij} := \frac{D_{ij}D_{N+1}}{D_jD_{i,N+1}} \quad , \quad 1 \leq i < j \leq N+1 \quad ,$$

as in Paper I, this yields

$$A_{n+1} = -\beta_{n,n+1}U_nA_n - v_n^2A_{n-1} + (1+v_n^2)A_n = T_nA_n - v_n^2A_{n-1} \quad , \quad (4.35)$$

where the $2\times 2$-matrices $T_n$ are defined as $T_n := (1+v_n^2)\mathcal{I} - \beta_{n,n+1}U_n$, $1 \leq n \leq N$ and the starting condition is $A_1 = \mathcal{I}$. This is the same recurrence relation as that in gravitational lens theory, see e.g. eq. (2.21) of Paper I. Hence we have shown that the recurrence relation for the mapping of the Jacobi matrices in lens theory can be derived as a direct approximation from geometrical optics.

### 4.4 The deflection angle, the lens equation

We have seen that light propagation for infinitesimal light beams can be derived from geometrical optics. Can one also derive the lens equation and the deflection angle from geometrical optics? Yes, provided that the matter outside the clumps is homogeneous and the source of shear due to the clumps is assumed to vanish outside of the clumps. Therefore, the mapping between two consecutive lens planes can be considered to be linear on a large scale, i.e., not just for infinitesimal beams, but also for 'fat beams' (which of course have to be smaller than the typical separation between clumps).

Consider two rays $\gamma_0$ and $\gamma_I$ including an angle $\boldsymbol{\theta}$ at their intersection point at the observer, where this angle is small enough to ensure that these rays are approximately parallel, but not necessarily infinitesimally small.

We treat one of them ($\gamma_0$) as a reference ray, adapt a screen to it (as defined in Sect. 2) and denote the screen position of $\gamma_I$ at redshift $z$ by $\boldsymbol{\xi}^I(z)$. We calculate the evolution of this separation vector from the observer ($z=0$) to the source at $z_s = z_{N+1}$ in two steps:

1) Due to the remark above, the separation vector has to satisfy the Jacobi deviation equation (2.2a) with the source term $\mathcal{T} = \mathcal{R}_{bg}\mathcal{I}$ outside inhomogeneities. Hence, each component of this separation vector has to satisfy the Dyer–Roeder differential equation (4.27). Thus, if we indicate the screen position of $\gamma_I$ (relative to $\gamma_0$) at the $j$-th inhomogeneity by $\boldsymbol{\xi}^I_j$ we can describe this separation vector between the $(n-1)$-th and the $n$-th lens plane by

$$\boldsymbol{\xi}^I(z) = \frac{r(z_n, z)}{r(z_n, z_{n-1})} \boldsymbol{\xi}^I_{n-1} + \frac{r(z_{n-1}, z)}{r(z_{n-1}, z_n)} \boldsymbol{\xi}^I_n \quad , \quad z_{n-1} \leq z \leq z_n \quad . \tag{4.36}$$

Note, that $r(z_n, z)$ and $r(z_{n-1}, z)$ form a pair of linearly independent solutions of the Dyer-Roeder equation and that inserting $z_n$ and $z_{n-1}$ yields the correct boundary conditions.

2) If there was no inhomogeneity at redshift $z_n$, (4.36) would stay valid also for $z \geq z_n$. But since there is an inhomogeneity, we have to correct for this and we have to take into account that for $z > z_n$, the optical tidal matrix again becomes $\mathcal{T} = \mathcal{R}_{bg}\mathcal{I}$. The correction function has to be a solution $B(z)$ of the Dyer-Roeder equation. Thus we obtain

$$\boldsymbol{\xi}^I(z) = \frac{r(z_n, z)}{r(z_n, z_{n-1})} \boldsymbol{\xi}^I_{n-1} + \frac{r(z_{n-1}, z)}{r(z_{n-1}, z_n)} \boldsymbol{\xi}^I_n - B(z)\, \mathbf{c}_n(\boldsymbol{\xi}^I_n) \quad , \quad z_n \leq z \leq z_{n+1} \quad . \tag{4.37}$$

$\mathbf{c}_n$ is a non-zero vector quantity, therefore $B$ must vanish at $z_n$. We can choose the derivative of $B$ at $z_n$ such that

$$\frac{\mathrm{d}B(\lambda)}{\mathrm{d}\lambda}\Big|_{\lambda=\lambda_n} = (1+z_n) \quad . \tag{4.38}$$

holds, and thus $B(z) = D(z_n, z)$.

**The deflection angle**

We define the derivatives of the separation vector of the two rays with respect to the affine parameter, before and after the $n$-th inhomogeneity:

$$\dot{\boldsymbol{\xi}}^I_{n+} := \lim_{\Delta\lambda \searrow 0} \frac{\mathrm{d}}{\mathrm{d}\lambda} \boldsymbol{\xi}^I(\lambda_n + \Delta\lambda) \quad , \quad \dot{\boldsymbol{\xi}}^I_{n-} := \lim_{\Delta\lambda \searrow 0} \frac{\mathrm{d}}{\mathrm{d}\lambda} \boldsymbol{\xi}^I(\lambda_n - \Delta\lambda) \quad . \tag{4.39}$$

Since $\mathrm{d}D_{\text{proper}} = (1+z_n)\mathrm{d}\lambda$ for an observer at $z_n$,

$$\mathbf{e}_{\text{out}} = (1+z_n)^{-1} \dot{\boldsymbol{\xi}}^I_{n+} \quad \text{and} \quad \mathbf{e}_{\text{in}} = (1+z_n)^{-1} \dot{\boldsymbol{\xi}}^I_{n-} \tag{4.40}$$

are the angular directions of $\gamma_I$ relative to $\gamma_0$ before ($\mathbf{e}_{\text{in}}$), and after traversing the inhomogeneity ($\mathbf{e}_{\text{out}}$), respectively. We use (4.36), (4.37) and (4.38) to obtain

$$\dot{\boldsymbol{\xi}}^{\mathrm{I}}_{n+} - \dot{\boldsymbol{\xi}}^{\mathrm{I}}_{n-} = -\lim_{\Delta\lambda \searrow 0} \frac{\mathrm{d}}{\mathrm{d}\lambda} B(\lambda_n + \Delta\lambda)\mathbf{c}_n(\boldsymbol{\xi}^{\mathrm{I}}_n) = -(1+z_n)\mathbf{c}_n(\boldsymbol{\xi}^{\mathrm{I}}_n) \quad ; \tag{4.41}$$

with (4.40) this becomes

$$(\mathbf{e}_{\mathrm{out}} - \mathbf{e}_{\mathrm{in}}) = -\mathbf{c}_n(\boldsymbol{\xi}^{\mathrm{I}}_n) \quad . \tag{4.42}$$

Hence, $\mathbf{c}_n(\boldsymbol{\xi}^{\mathrm{I}}_n)$ is the difference of the deflection angles at the screen position $\boldsymbol{\xi}^{\mathrm{I}}_n$ and the reference ray position ($\boldsymbol{\xi}_n = \mathbf{0}$). We now calculate the value of the vector $\mathbf{c}_n(\boldsymbol{\xi}^{\mathrm{I}}_n)$ as a function of the surface mass density $\Sigma$ of the inhomogeneity and show that it is equal to the difference of the deflection angles $\hat{\boldsymbol{\alpha}}_n(\boldsymbol{\xi}^{\mathrm{I}}_n) - \hat{\boldsymbol{\alpha}}_n(\mathbf{0})$ used in lens theory.

Consider a family of rays forming an infinitesimal beam with central ray $\gamma_{\mathrm{I}}$; we denote their screen vectors in the $n$-the lens plane by $\boldsymbol{\xi}_n = \boldsymbol{\xi}^{\mathrm{I}}_n + \Delta\boldsymbol{\xi}_n$ and their angular positions relative to $\gamma_{\mathrm{I}}$ at the observer by $\Delta\boldsymbol{\theta}$. Discussing the Jacobian map of this infinitesimal beam $\mathcal{D}_n(\boldsymbol{\xi}^{\mathrm{I}}_n) = \frac{\partial \Delta\boldsymbol{\xi}_n}{\partial \Delta\boldsymbol{\theta}}$ and its derivatives

$$\dot{\mathcal{D}}^+_n(\boldsymbol{\xi}^{\mathrm{I}}_n) = \frac{\partial \Delta\dot{\boldsymbol{\xi}}_{n+}}{\partial \Delta\boldsymbol{\theta}} \quad , \quad \dot{\mathcal{D}}^-_n(\boldsymbol{\xi}^{\mathrm{I}}_n) = \frac{\partial \Delta\dot{\boldsymbol{\xi}}_{n-}}{\partial \Delta\boldsymbol{\theta}} \quad ,$$

at the inhomogeneity, we obtain with (4.41) for the difference of these matrices

$$\dot{\mathcal{D}}^+_n(\boldsymbol{\xi}^{\mathrm{I}}_n) - \dot{\mathcal{D}}^-_n(\boldsymbol{\xi}^{\mathrm{I}}_n) = -(1+z_n)\left[\frac{\partial \mathbf{c}_n(\boldsymbol{\xi}_n)}{\partial \Delta\boldsymbol{\theta}}\right]\bigg|_{\boldsymbol{\xi}_n = \boldsymbol{\xi}^{\mathrm{I}}_n} =$$
$$= -(1+z_n)\left[\left(\frac{\partial \mathbf{c}_n(\boldsymbol{\xi}_n)}{\partial \boldsymbol{\xi}_n}\right)\left(\frac{\partial \boldsymbol{\xi}_n}{\partial \Delta\boldsymbol{\theta}}\right)\right]\bigg|_{\boldsymbol{\xi}_n = \boldsymbol{\xi}^{\mathrm{I}}_n} = -(1+z_n)\left[\left(\frac{\partial \mathbf{c}_n}{\partial \boldsymbol{\xi}_n}\right)\mathcal{D}_n(\boldsymbol{\xi}_n)\right]\bigg|_{\boldsymbol{\xi}_n = \boldsymbol{\xi}^{\mathrm{I}}_n} . \tag{4.43}$$

On the other hand, we have from (4.26)

$$\dot{\mathcal{D}}^+_n(\boldsymbol{\xi}^{\mathrm{I}}_n) - \dot{\mathcal{D}}^-_n(\boldsymbol{\xi}^{\mathrm{I}}_n) = -(1+z_n)\left[\tilde{U}_n(\boldsymbol{\xi}_n)\mathcal{D}_n(\boldsymbol{\xi}_n)\right]\bigg|_{\boldsymbol{\xi}_n = \boldsymbol{\xi}^{\mathrm{I}}_n} \quad .$$

This implies

$$\left[\frac{\partial \mathbf{c}_n(\boldsymbol{\xi}_n)}{\partial \boldsymbol{\xi}_n}\right]\bigg|_{\boldsymbol{\xi}_n = \boldsymbol{\xi}^{\mathrm{I}}_n} = \left[\tilde{U}_n(\boldsymbol{\xi}_n)\right]\bigg|_{\boldsymbol{\xi}_n = \boldsymbol{\xi}^{\mathrm{I}}_n} = \mathcal{H}\left[\tilde{\Psi}_n(\boldsymbol{\xi}_n)\right]\bigg|_{\boldsymbol{\xi}_n = \boldsymbol{\xi}^{\mathrm{I}}_n} \quad , \tag{4.44}$$

for every $\boldsymbol{\xi}^{\mathrm{I}}_n$ and therefore,

$$\frac{\partial \mathbf{c}_n(\boldsymbol{\xi}_n)}{\partial \boldsymbol{\xi}_n} = \mathcal{H}\left[\tilde{\Psi}_n(\boldsymbol{\xi}_n)\right] \quad , \quad \Rightarrow \quad \mathbf{c}_n(\boldsymbol{\xi}_n) = \nabla_{\boldsymbol{\xi}_n}\tilde{\Psi}_n(\boldsymbol{\xi}_n) + \mathbf{const} \quad . \tag{4.45}$$

The additive constant has to be $-\nabla_{\boldsymbol{\xi}_n}\tilde{\Psi}_n(\mathbf{0})$; this can be obtained from the limit $\boldsymbol{\theta} \to \mathbf{0}$, i.e., the case where the ray considered coincides with the reference ray: for this ray $\boldsymbol{\xi}(z) \equiv \mathbf{0}$. Therefore, we finally obtain with (4.18)

$$\mathbf{c}_n(\boldsymbol{\xi}_n) = \frac{4G}{c^2}\int_{\mathbb{R}^2}\Sigma(\boldsymbol{\xi}')\left[\frac{(\boldsymbol{\xi}_n - \boldsymbol{\xi}')}{|\boldsymbol{\xi}_n - \boldsymbol{\xi}'|^2} - \frac{(\mathbf{0} - \boldsymbol{\xi}')}{|\mathbf{0} - \boldsymbol{\xi}'|^2}\right]\mathrm{d}^2\xi' \equiv \hat{\boldsymbol{\alpha}}_n(\boldsymbol{\xi}_n) - \hat{\boldsymbol{\alpha}}_n(\mathbf{0}) \quad ; \tag{4.46}$$

as claimed before this is the difference of the deflection angles of the ray considered and the reference ray.

**The lens equation**

Evaluating (4.37) at redshift $z_{n+1}$, inserting $B(z_{n+1}) = D(z_n, z_{n+1})$, (4.46), (4.29) with $r(z_n, z_{n-1}) = -|r(z_n, z_{n-1})|$, Etherington's reciprocity relation, and dropping the indices 'I' yields:

$$\boldsymbol{\xi}_{n+1} = -\frac{(1+z_{n-1})D_{n,n+1}}{(1+z_n)D_{n-1,n}} \boldsymbol{\xi}_{n-1} + \frac{D_{n-1,n+1}}{D_{n-1,n}} \boldsymbol{\xi}_n - D_{n,n+1} \left[\hat{\boldsymbol{\alpha}}_n(\boldsymbol{\xi}_n) - \hat{\boldsymbol{\alpha}}_n(\boldsymbol{0})\right] \quad . \quad (4.47)$$

Using the quantities $v_n^2$ and $\beta_{n,n+1}$ and the dimensionless impact vectors $\mathbf{x}_j = \boldsymbol{\xi}_j/D_j$ shows that the first term on the rhs of (4.47) can be rewritten as

$$-\frac{(1+z_{n-1})D_{n,n+1}}{(1+z_n)D_{n-1,n}} \boldsymbol{\xi}_{n-1} = -v_n^2 D_{n+1} \mathbf{x}_{n-1} \quad ; \quad (4.48a)$$

for the second one, using the equations (C2) and (C5) of Paper I, we obtain:

$$\frac{D_{n-1,n+1}}{D_{n-1,n}} \boldsymbol{\xi}_n = D_{n+1}(1+v_n^2) \mathbf{x}_n \quad . \quad (4.48b)$$

With the definition of the scaled deflection angle $\boldsymbol{\alpha} := \frac{D_{js}}{D_s} \hat{\boldsymbol{\alpha}}$, we find

$$D_{n,n+1} \left[\hat{\boldsymbol{\alpha}}_n(\mathbf{x}_n) - \hat{\boldsymbol{\alpha}}_n(\mathbf{0})\right] = D_{n+1} \beta_{n,n+1} \left[\boldsymbol{\alpha}_n(\mathbf{x}_n) - \boldsymbol{\alpha}_n(\mathbf{0})\right] \quad ; \quad (4.48c)$$

inserting the equations (4.48) in (4.47) yields the dimensionless recurrence relation for the impact vectors $\mathbf{x}_j$ in the lens planes

$$\mathbf{x}_{n+1} = (1+v_n^2) \mathbf{x}_n - v_n^2 \mathbf{x}_{n-1} - \beta_{n,n+1} \left[\boldsymbol{\alpha}_n(\mathbf{x}_n) - \boldsymbol{\alpha}_n(\mathbf{0})\right] \quad , \quad 1 \le n \le N \quad . \quad (4.49)$$

We transform the center of the coordinate system in each lens plane such that

$$\mathbf{x}'_j := \mathbf{x}_j - \sum_{i=1}^{j-1} \beta_{ij} \boldsymbol{\alpha}_i(\mathbf{0}) \quad , \quad (4.50a)$$

define

$$\boldsymbol{\alpha}'_j(\mathbf{x}'_j) := \boldsymbol{\alpha}_j(\mathbf{x}_j) \quad (4.50b)$$

and obtain with (C8) of Paper I and the comment below this equation in Paper I, the recurrence relation one uses in lens theory [see Paper I, equation (2.19)]:

$$\mathbf{x}'_{n+1} = (1+v_n^2) \mathbf{x}'_n - v_n^2 \mathbf{x}'_{n-1} - \beta_{n,n+1} \boldsymbol{\alpha}'_n(\mathbf{x}'_n) \quad . \quad 1 \le n \le N \quad . \quad (4.51)$$

Whereas (4.49) describes the mapping of a ray relative to a reference ray, which is also deflected at every inhomogeneity, (4.51) describes the mapping of a ray relative to the 'optical axis'. This optical axis can be constructed by piecewise smooth null geodesics (of the empty cone metric) connecting the (new) centers of the coordinate systems on consecutive lens planes with each other; thus this optical axis represents a kinematically possible ray, but not necessarily an actual light ray (see Fermats principle in SEF, e.g., Chapt. 9.2). It has been shown already in SEF that the formulation (4.51) of the multiple lens plane equation is equivalent to the more familiar one (now we drop the primes),

$$\mathbf{x}_j = \mathbf{x}_1 - \sum_{i=1}^{j-1} \beta_{ij} \boldsymbol{\alpha}_i(\mathbf{x}_i) \quad , \quad 1 \le j \le N+1 \quad ,$$

for the special case $j = N+1$, we obtain with $\beta_{i,N+1} = 1$ for the source position that

$$\mathbf{y} := \mathbf{x}_{N+1} = \mathbf{x}_1 - \sum_{i=1}^{N} \boldsymbol{\alpha}_i(\mathbf{x}_i) \quad .$$

Therefore, we have shown in this chapter that the equations describing the mapping of a light ray and that of a light beam in gravitational lens theory can be derived with a series of well defined approximations from the description of light propagation in geometrical optics. In essence, the multiple deflection gravitational lens equation can be viewed as a discretization of the exact propagation equation (2.4), applied to the case of weak gravitational fields (but not necessarily weak matter inhomogeneities).

### 4.5 Remark on Fermat's principle

In SEF, Sect. 4.6, the derivation of the lens equation was based on a relativistic version of Fermat's principle. The argument leading to the geometric contribution to the time delay, eq. (4.65), p. 145 in SEF, suffers from an apparent inconsistency. On p. 143, it is first stated that light rays from the source to the neighborhood of the deflector and, after deflection, those from that neighborhood to the observer, form 'shearfree beams … subject only to the focussing of the smooth part of matter', i.e. to $\tilde{\alpha}\rho_F$; but the subsequent calculations are said to be based on the large-scale RW metric which is related to the average density, $\rho_F$. This, however, presents an apparent difficulty only. In the 'empty' region, outside clumps, the shear of light beams is assumed negligible there. Now, it is known that the only conformally flat non-static dust spacetimes are Friedmann ones (see Kramer et al. 1980, Sects. 22.2, 32.42, 32.5). Therefore, it seems reasonable to approximate the universe in 'empty' cone regions by a Friedmann model whose mean motion equals that of the large-scale background model, but whose density is $\tilde{\alpha}\rho_F$. This implies that the metric, $\bar{\mathrm{d}s}^2$, is related to the large-scale metric, $\mathrm{d}s^2$, by a constant conformal factor,

$$\bar{\mathrm{d}s}^2 = \tilde{\alpha}^{-1} \mathrm{d}s^2 = \tilde{\alpha}^{-1} R^2(\eta) \left\{ \mathrm{d}\eta^2 - \mathrm{d}\sigma_k^2 \right\} \quad .$$

Therefore, the spatial paths of light rays in empty regions are the same for $\bar{\mathrm{d}s}^2$ as for $\mathrm{d}s^2$, viz. geodesics 'of $\mathrm{d}\sigma_k^2$', and the reasoning on p. 144/145 leading to eq. (4.65) applies without change, since that equation is invariant under a constant rescaling of the RW metric. (Angles and the redshift $z_d$ remain unchanged, and the distances $c\Delta_{\mathrm{geom.}}$, $D_d$, $D_s$, $D_{ds}$ are rescaled by the same factor.)

## 5 The magnification of the flux of light beams

### 5.1 The flux of a radiation field, magnification factor

The monochromatic flux $S_\omega$ of a radiation field, measured by an observer at frequency $\omega$, is given by the product of its specific intensity $I_\omega$ and the solid angle $\mathrm{d}\Omega$ the source subtends on the observers sky: $S_\omega = I_\omega \mathrm{d}\Omega$. The specific intensity at the observer is related to that at the source by the conservation of the phase space density of photons.

This implies, according to SEF, Sect. 3.6, that for any non-interacting radiation field the scalar $\frac{I_\omega}{\omega^3}$ is observer-independent, i.e., independent of his four velocity, and constant on a light beam:

$$\frac{I_{\omega(\lambda)}(\lambda)}{\omega^3(\lambda)} = \frac{I_{\omega_s}(\lambda_s)}{\omega_s^3} = \frac{I_{\omega_0}(0)}{\omega_0^3} \quad , \tag{5.1}$$

where $\lambda$ is the affine parameter of the central light ray of the beam, $\omega(0) =: \omega_0$ and $\omega(\lambda_s) =: \omega_s$.

Consider an infinitesimal monochromatic source radiating with frequency $\omega_s$, and observed with frequency $\omega$; its observed flux $S_\omega$ depends on the source of shear and convergence along the beam connecting source and observer. Changing these source terms such that the frequency at the observer and the affine parameter-redshift relation stays the same, then, for the same observer, the observed flux of the source changes according to (5.1) to $S_\omega = S_\omega^0 \frac{d\Omega}{d\Omega^0}$, with $S_\omega^0$ being the flux before changing the source terms. In an on-average Friedmann universe, the frequency of the light is not changed by the deflection and, by definition, the affine parameter-redshift relation is not affected by the clumps. Hence, we can compare the flux $S_\omega$ of the source with the case where there are no intervening clumps between source and observer, and obtain for the ratio

$$\mu := \frac{S_\omega}{S_\omega^0} = \frac{d\Omega}{d\Omega^0} \quad , \quad 0 \leq \mu \quad . \tag{5.2}$$

$\mu$ is the so-called *magnification factor*; if $\mu > 1$, the light beam is called *magnified* relative to the empty beam. $d\Omega$ and $d\Omega^0$ are the solid angles which the source subtends on the sky for the cases with and without clumps in the beam. If we use (2.7), we obtain that the magnification $\mu(\lambda)$ of a source at the affine parameter $\lambda$ compared to the case where the source is observed through the empty beam, can be described as

$$\mu(\lambda) = \left| \frac{\det \mathcal{D}^0(\lambda)}{\det \mathcal{D}(\lambda)} \right| = \left| \frac{D^2(\lambda)}{\det \mathcal{D}(\lambda)} \right| = \left| \frac{1}{\det A(\lambda)} \right| \quad . \tag{5.3}$$

For the second equality we have used that for the empty beam, the Jacobi matrix is given by $\mathcal{D}^0(\lambda) = D(\lambda)\mathcal{I}$, with $D(\lambda)$ being the angular diameter distance of the empty beam, i.e., the solution of the Dyer-Roeder equation (4.27) with boundary conditions (4.28). The third equality follows from the definition of the dimensionless Jacobian matrix $A(\lambda) = \frac{1}{D(\lambda)}\mathcal{D}(\lambda)$. Hence, the discussion of the matrix $A$ or the magnification factor $\mu$ in gravitational lens theory always implies the discussion of light propagation relative to the empty beam case. This point of view is reasonable:

1) As long as there are only a few clumps, i.e, if $1 - \tilde{\alpha}$ is small, most light beams are empty cone beams. Therefore, the magnification factor in (5.3) describes the observed flux of a source whose beam is distorted between source and observer, relative to the most typical case, where the beam is not distorted.

2) The other extreme is the case where $1 - \tilde{\alpha}$ becomes approximately one: the source of convergence becomes extremely small, and for the description of the very few light beams that do not traverse a matter inhomogeneity, one cannot neglect the source of shear, which is different along every individual beam. Hence, there does no longer exist a typical light beam, and the definition of the magnification factor as in (5.2) and (5.3) has no illustrative meaning: it compares the flux of the considered light beam with that of a ficticious beam.

3) As mentioned at the beginning of Sect. 4, for a weakly inhomogeneous universe (e.g., if one considers spatial scales on which the matter inhomogeneities are (quasi-)linear), the magnification is defined relative to that of the smooth Friedmann-Lemaître universe. In this case, the angular diameter distances $D(\lambda)$ are those obtained from (4.27) with $\tilde{\alpha} = 1$, and $\rho_{\rm cl} = \delta\rho$ (the density fluctuations) can have either sign – therefore, $\mathcal{R}_{\rm cl}$ no longer is non-positive.

## 5.2 The relative focusing equation

The focusing equation (2.8) describes the evolution of the angular diameter distance of a light beam due to the Ricci-focusing and the shear rate of the beam. In the case of an on-average Friedmann universe, all light beams have the empty cone background density as a common contribution to their focusing, and different additional source terms due to the clumps. Therefore, we want to derive a differential equation which describes the evolution of the beam relative to the empty beam; the source terms of this relative focusing equation are then produced by the clumps only.

Consider the differential equation

$$\frac{\mathrm{d}^2}{\mathrm{d}\lambda^2} w(\lambda) = [h(\lambda) + c(\lambda)]\, w(\lambda) \quad, \tag{5.4}$$

and let $w(\lambda)$ be the (unknown) solution of (5.4) with boundary conditions $w(0) = 0$ and $\dot{w}(0) = 1$. Assume that $v(\lambda)$ is the well-known solution of (5.4) for the case $c(\lambda) \equiv 0$,

$$\frac{\mathrm{d}^2}{\mathrm{d}\lambda^2} v(\lambda) = h(\lambda) v(\lambda) \quad, \tag{5.5}$$

with the same boundary conditions: $v(0) = 0$, $\dot{v}(0) = 1$. We define a strictly monotonically decreasing function $X(\lambda)$ by

$$X(\lambda) := \int_{\lambda}^{\lambda_{\max}} \frac{\mathrm{d}\lambda}{v^2(\lambda)} \quad, \tag{5.6}$$

so that $X(\lambda_{\max}) = 0$; the value of $\lambda_{\max}$ will be specified below. Then, inserting the equations (5.5) and (5.6) in (5.4), we obtain for the ratio $a := \frac{w}{v}$ the differential equation:

$$\frac{\mathrm{d}^2}{\mathrm{d}X^2}\, a(X) = v^4(X)\, c(X)\, a(X) \quad; \tag{5.7}$$

Using $\ddot{w}(\lambda)|_{\lambda=0} = 0 = \ddot{v}(\lambda)|_{\lambda=0}$, the boundary conditions for $a$ become, as a function of $\lambda$,

$$a(\lambda)|_{\lambda=0} = 1 \quad, \quad \frac{\mathrm{d}}{\mathrm{d}\lambda} a(\lambda)|_{\lambda=0} = 0 \quad. \tag{5.8}$$

We interprete[8] (5.4) by inserting $h = \mathcal{R}_{\rm bg}(\lambda)$ and $c = \mathcal{R}_{\rm cl}(\lambda) - |\sigma(\lambda)|^2$; then, $w$ and $v$ denote the angular diameter distances of the 'actual' beam considered and that in an empty cone, respectively. Therefore, (5.7) describes how the considered light beam is

---
[8] One can calculate the relative magnification of two light beams with (5.7) even in a case of a non Friedmann universe, if the affine parameters of these light beams are the same (e.g. as a function of redshift).

focused relative to the empty beam and is therefore called *relative focusing equation*; the solution of (5.7) can be described, with

$$v(\lambda) = D(\lambda) = \frac{c}{H_0} r(z(\lambda)) \quad ,$$

as

$$a(\lambda) = \frac{1}{D(\lambda)} \mathcal{SQ} \left[ \det \mathcal{D} \right] (\lambda) = \mathcal{SQ} \left[ \det A(\lambda) \right] \quad . \tag{5.9}$$

The inverse of (5.9) yields the magnification of the beam at a position $\lambda$: $|a(\lambda)|^2 = \mu^{-1}(\lambda)$. We can identify $\frac{c}{H_0} X$ with the cosmological $\chi$-function, defined in equation (4.68) of SEF, since

$$\frac{\mathrm{d}X}{\mathrm{d}\lambda} = -\left(\frac{c}{H_0}\right)^{-2} \frac{1}{r^2(\lambda)}$$

yields, if we put $\lambda_{\max} = \lim_{z \to \infty} \lambda(z)$ and use equation (4.3),

$$X(z) = \left(\frac{c}{H_0}\right)^{-1} \int_z^\infty \frac{\mathrm{d}z'}{r^2(z')(1+z')^3 \sqrt{\Omega z' + 1}} = \left(\frac{c}{H_0}\right)^{-1} \chi(z) \quad . \tag{5.10}$$

Inserting (5.10) and (5.9) in (5.7), the relative focusing equation can be rewritten as

$$\frac{\mathrm{d}^2}{\mathrm{d}\chi^2} a(\chi) = \left(\frac{c}{H_0}\right)^2 r^4(\chi) \left[ \mathcal{R}_{\mathrm{cl}} - |\sigma(\chi)|^2 \right] a(\chi) \quad , \quad a(\chi) = \mathcal{SQ} \left[ \det A \right] (\chi) \quad . \tag{5.11}$$

Note, that due to the strictly monotonic behaviour of $\chi$ and $\lambda$ as functions of $z$, we can consider any variable on a light ray as a function of $z$, $\lambda$ or $\chi$.

### 5.3 The focusing theorem

The non-positiveness of the source term $\mathcal{R}_{\mathrm{cl}} - |\sigma|^2$ due to the clumps in the focusing equation shows that a beam propagating through clumps is always more focused than the empty beam (in the absence of conjugate points between source and observer). Hence, as long as the beam has not formed its first conjugate point, the angular diameter distance must not be greater than that of an empty comparison beam at the same redshift. This so-called *focusing theorem* can be restated with the use of the relative focusing equation: As long as the light beam has not formed its first conjugate point, the function $a(\lambda)$ is alway between one and zero,

$$1 \geq a(\lambda) \geq 0 \iff \mu(\lambda) \geq 1 \quad , \quad 0 \leq \lambda \leq \lambda_{\mathrm{c}} \quad , \tag{5.12}$$

or, the light beam is not demagnified relative to one in an empty cone. This can be proven immediately: Using the boundary conditions of $a(\lambda)$ described in equation (5.8) and that, due to the non-positiveness of $\left[ \mathcal{R}_{\mathrm{cl}} - |\sigma|^2 \right]$, the second derivative of $a$ in (5.11) is always non-positive, one obtains that the value of $a$ is always between one and zero in the interval between the vertex and the first conjugate point of the beam. One can also prove a stronger statement: *as long as the beam has not passed a conjugate point, the function $a(\lambda)$ is monotonically decreasing.*
Proof: Since $\chi$ tends to plus infinity at the vertex, and $\lim_{\chi \to \infty} \frac{\mathrm{d}a}{\mathrm{d}\chi}(\chi) = 0$, one can write

$$\frac{\mathrm{d}a}{\mathrm{d}\chi}(\chi) = \int_{\infty}^{\chi} \left(\frac{c}{H_0}\right)^2 r^4(\chi') \left[\mathcal{R}_{\mathrm{cl}} - |\sigma|^2\right] a(\chi') \,\mathrm{d}\chi' \quad ,$$

and therefore, $\frac{\mathrm{d}a}{\mathrm{d}\lambda}$ can be rewritten with as

$$\frac{\mathrm{d}a}{\mathrm{d}\lambda}(\lambda) = \left(\frac{c}{H_0}\right) \frac{1}{r^2(\lambda)} \int_{\chi(\lambda)}^{\infty} r^4(\chi') \left[\mathcal{R}_{\mathrm{cl}}(\chi') - |\sigma|^2(\chi')\right] a(\chi') \,\mathrm{d}\chi' \quad .$$

Since the integrand is non-positive, $\frac{\mathrm{d}a}{\mathrm{d}\lambda} \leq 0$ follows. q.e.d.

# 6 Summary and conclusions

We have investigated the propagation of infinitesimally small light beams in arbitrary spacetimes and derived a Jacobi-type differential equation for the matrix providing the linear mapping from the inclination angle of a light ray of the beam to the separation vector at arbitrary values of the affine parameter. This matrix carries full information about the size, shape, orientation and twist of the beam. We have then concentrated on the investigation of the behaviour of light beams near a vertex and near conjugate points; in particular, we have derived asymptotic representations of the optical scalars near such points. It was pointed out that near a vertex and a focus, the twist of a beam is a higher-order contribution to the Jacobi mapping than are expansion and shear.

We then turned to the special case that the metric is that of a perturbed Friedmann universe, i.e., where the overall geometry of the universe is described by a Friedmann metric, which however is locally modified to allow for matter inhomogeneities. If the matter inhomogeneities are considered to be weak, so that they can be described locally by a post-Minkowskian metric, and geometrically-thin and isolated, so that typical light beams are propagating most of the time through the background Friedmann metric, the influence of the matter inhomogeneities on the light beam can be described by a sum of delta-distributional contributions to the source term of the Jacobi equation for the linear mapping mentioned above. In this way, we have derived the equations of gravitational lens theory, which represents an approximation to the exact propagation equations which is particularly useful for, and applies to, most astrophysically relevant situations of light propagation in the universe. We want to point out that in contrast to earlier treatments of the lens equations (e.g., SEF, Sect. 4.6), we have made no use of the existence of an optical axis relative to which the impact vectors are defined; instead, our reference ray is a physical, i.e., deflected, light ray. To relate our formulation to the earlier treatment, a redefinition of the coordinate frames in the lens planes was performed which yielded the lens equation in the standard form. We remind the reader that a derivation of the gravitational lens equation can also start from Fermat's principle (see Blandford & Narayan 1986, SEF, Sect. 4.6 and references therein); however, the derivation presented here appears to be more dircet in that one does not make use of geometrical constructions for the calculation of the 'geometrical time delay', which are less easy to justify in an 'on-average-Friemann-universe' than the approximations used here. The advantage of our derivation of the lens equations lies in its explicit listing of approximations which have to be made. All but two are not critical and well satisfied in astrophysically relevant situations. The two which are as yet not very well understood are: (1) The source of shear was assumed to vanish between two consecutive lens planes. (2) It was assumed that the metric of a clumpy universe can be written locally as a post-Minkowskian modification of the standard Friedmann metric. Note that a number of investigations have suggested the validity of this latter approximation (e.g., Futamase & Sasaki 1989, Jacobs et al. 1993). The former assumption certainly has to break down if the universe is highly clumpy, i.e., for $\tilde{\alpha}\Omega$ of order unity. However, since it seems that the clumpiness of our universe is much smaller than unity, we conclude that the (multiple deflection) gravitational lens equations provide a useful and fairly accurate approximation in most relevant cases.

Finally, we have derived an equation for the size of a light beam in a clumpy universe, relative to the size of a beam which is unaffected by the matter inhomogeneities. If we require that this second-order differential equation contains only the contribution by matter clumps as source term, the independent variable is uniquely defined and agrees with the $\chi$-function previously introduced [see SEF, eq. (4.68)] for other reasons. This relative focusing equation immediately yields the result that a light beam cannot be less focused than a reference beam which is unaffected by matter inhomogeneities, prior to the propagation through its first conjugate point. In other words, no source can appear fainter to the observer than in the case that there are no matter inhomogeneities close to the line-of-sight to this source, a result previously demonstrated for the case of one (Schneider 1984) and several (Paper I, Seitz & Schneider 1994) lens planes.